\documentclass[article]{jsedi_sub}
\usepackage[caption=false]{subfig}

\usepackage{epstopdf}
\makeatletter
\newcommand{\raisemath}[1]{\mathpalette{\raisem@th{#1}}}
\newcommand{\raisem@th}[3]{\raisebox{#1}{$#2#3$}}
\makeatother
\newcommand{\tavg}[1]{\overline{\raisebox{0pt}[1.1\height]{$\mkern+4mu #1 \mkern+4mu$}}^{\raisemath{-0.5pt}{\, t}}}

\def\be{\begin{equation}}
\def\ee{\end{equation}}

\def\begineqn{\begin{equation*}}
\def\endeqn{\end{equation*}}

\def\beginar{\begin{eqnarray}}
\def\endar{\end{eqnarray}}
\def\beginarn{\begin{eqnarray*}}
\def\endarn{\end{eqnarray*}}

\def\lb{\left ( }
\def\rb{\right ) }

\def\ub{\mathbf{u}}

\def\ubp{\mathbf{u}^{\prime}}

\def\dst{{\partial_t}}

\def\hz{{\bf\widehat z}}

\def\hr{{\bf \widehat r}}

\def\Rat{\widetilde{Ra}}

\title{Asymptotics of spherical dynamos exhibiting a small-scale MAC balance}
\shorttitle{Asymptotics of spherical dynamos} 

\author[1]{Justin A. Nicoski
	\orcid{0009-0002-3065-100X}
}
\author[1]{Andy Esseln
	\orcid{0009-0002-4493-7797}
}
\author[1]{Michael A. Calkins
	\orcid{0000-0002-6917-5365}
		\thanks{Corresponding author: 
\href{mailto:michael.calkins@colorado.edu}{\texttt{michael.calkins@colorado.edu}}}
}
\author[2]{Chris Davies
	\orcid{0000-0002-1074-3815}
}
\affil[1]{Department of Physics, University of Colorado Boulder, Boulder, Colorado USA}
\affil[2]{School of Earth and Environment, University of Leeds, Leeds LS2 9JT, United Kingdom }

\credit{Conceptualization}{Michael A. Calkins}
\credit{Data curation}{Justin A. Nicoski, Andy Esseln, Michael A. Calkins}
\credit{Formal analysis}{Justin A. Nicoski, Andy Esseln, Michael A. Calkins, Chris Davies}
\credit{Funding acquisition}{Michael A. Calkins}
\credit{Investigation}{Justin A. Nicoski, Andy Esseln, Michael A. Calkins, Chris Davies}
\credit{Writing - Original draft}{Justin A. Nicoski, Michael A. Calkins, Chris Davies}
\credit{Writing - Review \& Editing}{Justin A. Nicoski, Michael A. Calkins, Chris Davies}

\begin{document}
%\maketitle

\makesedititle{
  \begin{summary}{Abstract}
Understanding the asymptotic behaviour of numerical dynamo models is critical for extrapolating results to the physical conditions that characterise terrestrial planetary cores.
Here we investigate the behaviour of convection-driven dynamos reaching a MAC (magnetic-Archimedes-Coriolis) balance on the convective length scale and compare the results with non-magnetic convection cases. 
In particular, the dependence of physical quantities on the Ekman number, $Ek$, is studied in detail. 
The scaling of velocity dependent quantities is observed to be independent of the force balance and in agreement with quasi-geostrophic theory. The primary difference between dynamo and non-magnetic cases is that the fluctuating temperature is order unity in the former such that the buoyancy force scales with the Coriolis force. The MAC state yields a scaling for the flow speeds that is identical to the so-called CIA (Coriolis-inertia-Archimedes) scaling.
There is an $O(Ek^{1/3})$ length scale present within the velocity field irrespective of the leading order force balance. This length scale is consistent with the asymptotic scaling of the terms of the governing equations and is not an indication that viscosity plays a dominant role. The peak of the kinetic energy spectrum and the ohmic dissipation length scale both exhibit an Ekman number dependence of approximately $Ek^{1/6}$, which is consistent with a scaling of $Rm^{-1/2}$, where $Rm$ is the magnetic Reynolds number. For the dynamos, advection remains comparable to, and scales similarly with, both inertia and viscosity, implying that nonlinear convective Rossby waves play an important role in the dynamics even in a MAC regime.
    \vspace{1.3cm} % because abstract is too short
  \end{summary}
% \begin{summary}{Non-technical summary}
%    The text goes here. Again, no longer than 200 words, no reference.
%    \vspace{1.3cm} % because abstract is too short
% \end{summary}
}

%\linenumbers

\section{Introduction}
\label{S:intro}

The magnetic fields of the Earth, planets and stars are all thought to be generated by the motion of electrically conducting fluids \citep[e.g.][]{cJ11b}. In the case of the Earth, this dynamo process occurs in the liquid iron outer core \citep{pR13}. The core is likely a turbulent environment and therefore characterised by a broad range of spatial and temporal scales \citep{jmA15}, which has necessitated the use of numerical models for studying its dynamics \citep[e.g.][]{kZ89,aK95,gG95}. These early simulations exhibited magnetic fields that show many similarities with the geomagnetic field, including a predominantly dipolar structure and the presence of polarity reversals \citep{gG95}. Since \cite{gG95}, many dynamo simulations have been performed which have qualitative similarities with the Earth's magnetic field \citep[e.g.][]{uC10b,dM21}. 
As such, dynamo simulations are believed to capture certain aspects of the dynamics of the geodynamo.

Due to their high computational cost, direct numerical simulations (DNS) cannot yet reach the parameter space of the geodynamo. In particular, the Ekman number, $Ek = \nu/ ( \Omega H^2 )$, characterising the ratio of the viscous force to the Coriolis force at large length scales, is approximately $Ek \sim 10^{-15}$ in the outer core (here $\nu$ is the kinematic viscosity, $\Omega$ is the angular frequency or rotation rate and $H$ is the depth of the fluid layer). In contrast, DNS of spherical dynamos rarely reach Ekman numbers smaller than $Ek \sim 10^{-7}$ \citep{nS17,yL25}. Another important nondimensional parameter is the magnetic Prandtl number, which is defined as $Pm=\nu/\eta$, where $\eta$ is the magnetic diffusivity. Studies indicate that $Pm = O(10^{-6})$ in the Earth \citep[e.g.][]{mP13}, which is also far beyond what can currently be investigated in DNS due to the large flow speeds necessary to sustain dynamo action in such a regime. 

There are three modeling strategies that can be used to help bridge this gap. Historically, the primary approach consists of carrying out DNS at ever more extreme parameters and developing scaling laws from the output such that predictions can be made in an Earth-like regime \citep[e.g.][]{uC06,pD13b}. Scaling laws exploit dominant balances in the governing equations and must make assumptions with regard to how the spatiotemporal dynamics depend on system parameters.
Another approach involves assuming that the small-scales are of negligible dynamical importance such that only the large scales in the system are directly modeled. Hyperviscous damping on the smallest simulated scales can be used for numerical stabilisation, allowing for considerable extension of the parameter space  \citep{jA17, cG19, jA23}.
A third approach is a model reduction strategy that involves identifying the small parameters of the system and using perturbation methods to formally derive a new, simpler, equation set that is significantly cheaper to solve, thereby allowing access to more extreme parameter space. This strategy has been successful in the planar \citep{kJ98a,mS06,kJ12,mC15b} and `annulus' models \citep{fB70,mC13} but has not been applied to the nonlinear spherical problem due in part to the variety of length scales present.  Towards this end, one of the main goals of the present work is to quantify the Ekman number dependence of the small-scale (i.e.~convection-scale) balances in the spherical geometry when strong magnetic field is present, which is a crucial aspect for understanding the asymptotic scaling of the system.

The derivation of scaling laws and asymptotic expansions relies on identification of the dominant force balance in the dynamics. Numerical simulations of rapidly rotating dynamos have revealed two main force balances, depending on the relative size of the Lorentz force, which is controlled primarily by the value of $Pm$. When the Lorentz force is sufficiently small, the leading order force balance is between the Coriolis force and the pressure gradient force and the resulting dynamics\textcolor{black}{, which arise from perturbations to this balance, are termed} quasi-geostrophic (QG) \citep{kS12,rY16,mC18,mY22,mY22b}. Some studies find that QG dynamos exhibit many quantitative similarities with non-magnetic QG convection \citep{kS12,mY22b}. In this sense, these simulations are similar to rapidly rotating convective cases without a magnetic field that develop a balance between the Coriolis and pressure forces at zeroth order, with the rest of the terms entering at first order, as described in asymptotic models \citep[e.g.][]{mS06}. Based on this line of reasoning, asymptotic dynamo models in planar geometries can be derived assuming that the Lorentz force enters at the same order as the first-order terms that appear in asymptotic quasi-geostrophic convection models (such as the viscous force, the advective term, and the buoyancy force) \citep{mC15b}. Simulations of rotating plane layer dynamos at small magnetic Prandtl numbers find the predicted asymptotic scalings from the quasi-geostrophic dynamo model, with the Lorentz force entering at the same order as the viscous force \citep{mY22}. Even within the QG limit, however, it is possible to generate magnetic field strengths such that the dynamics are vastly different relative to non-magnetic rotating convection \citep[e.g.][]{mP18,sM19}.

In contrast to the geostrophic balance, a MAC (magnetic-Archimedes-Coriolis) balance occurs when the Lorentz force is comparable in magnitude to the Coriolis and buoyancy forces \citep{pR13}. It has long been suggested that the Earth might be in a such a regime \citep{jT63}, and numerical simulations also suggest this possibility can arise on certain length scales \citep[e.g.][]{kS15,eD16,jmA17}. Numerical simulations can only achieve this regime with unrealistically large values of $Pm$ \citep[e.g.][]{eD16, lP18, mM20}. However, such studies are nevertheless important for understanding the differences between QG and MAC dynamos. On the other hand, scale-dependent force balances often suggest that the primary force balance is QG at large length scales, so it is sometimes argued that dynamos are QG-MAC rather than simply MAC, at least at large length scales \citep{jA17,tS19,tS21}.

Given that the force balance can be very different for dynamo simulations with a strong magnetic field as compared to convective cases without a magnetic field, the question arises as to how the magnetic field might alter the asymptotic relations found in rotating convection. We study this question by comparing the dynamo cases from \cite{mC21} with non-magnetic rotating convection cases and comparing the asymptotic behaviour of the two sets of data. The dynamo cases from \cite{mC21} are run at a fixed $Pm=2$, and develop a strong magnetic field, especially as the Ekman number is reduced. For many of our cases, the fluctuating Lorentz force is over half as large as the fluctuating Coriolis force, which ensures that our dynamo simulations are strongly influenced by the magnetic field. The paper is structured as follows. In section \ref{S:model}, the model equations, numerical method, and outputs are described. Section \ref{S:qg_scalings} briefly summarizes the expected asymptotic scalings of flow speeds, length scales, and forces in non-magnetic rapidly rotating convection. Section \ref{S:results} compares the asymptotic scalings between the non-magnetic and dynamo cases. Finally, a discussion of the results is provided in section \ref{S:conclusion}.

\section{Model}
\label{S:model}
We consider a self-gravitating Boussinesq fluid contained between rotating spherical shells. The radii of the inner and outer shell are denoted by $r_i$ and $r_o$, respectively. The aspect ratio is fixed for all cases to be $r_i/r_o = 0.35$. The inner and outer boundaries are held at constant temperatures $T_i$ and $T_o$, respectively, where $\Delta T = T_i - T_o > 0$. Using spherical coordinates $(r,\theta,\phi)$, the non-dimensional governing equations are given by
\be
\begin{aligned}
\dst \ub + \ub \cdot \nabla \ub = &-\frac{2}{Ek} \hz \times \ub - \frac{1}{Ek}\nabla P + \frac{Ra}{Pr} \lb \frac{r}{r_o} \rb T \, \widehat{\mathbf{r}}\\ &+ \frac{1}{EkPm}\mathbf{J}\times\mathbf{B}+ \nabla^2 \ub,
\end{aligned}
\ee
\be
\partial_t \mathbf{B} = \nabla \times \left(\ub \times \mathbf{B}\right) + \frac{1}{Pm} \nabla^2 \mathbf{B},
\ee
\be
\dst T + \ub \cdot \nabla T = \frac{1}{Pr} \nabla^2 T,
\ee
\be
\nabla \cdot \ub = 0, \qquad \nabla \cdot \mathbf{B} = 0,
\ee
where $\ub=\lb u_r, u_\theta, u_\phi \rb$ is the fluid velocity, $\mathbf{B}=\lb B_r, B_\theta, B_\phi \rb$ is the magnetic field, $\mathbf{J}=\nabla\times\mathbf{B}$ is the current density, $T$ is the temperature, $P$ is the pressure, $\widehat{\mathbf{r}}$ is a unit vector pointing in the outward radial direction, and $\hz$ is a unit vector that points parallel to the rotation axis.
The equations have been non-dimensionalised with shell depth $H = r_o - r_i$, viscous diffusion time $H^2/\nu$, temperature scale $\Delta T$, and magnetic field scale $\sqrt{\rho \mu \eta \Omega}$, where $\rho$ is the fluid density and $\mu$ is the vacuum permeability. 
The non-dimensional control parameters are given by
 \be
Ra=\frac{g_o\alpha \Delta T H^3}{\nu \kappa}, \quad Ek=\frac{\nu}{\Omega H^2}, \quad Pr = \frac{\nu}{\kappa}, \quad Pm=\frac{\nu}{\eta} .
\ee
Here, $Ra$ is the Rayleigh number, $Pr$ is the Prandtl number, $g_o$ is the acceleration due to gravity at the outer boundary, $\alpha$ is the thermal expansion coefficient, $\kappa$ is the thermal diffusivity, and $\eta$ is the magnetic diffusivity. For our simulations, we fix $Pr=1$ and most of the dynamo simulations use $Pm=2$. A subset of simulations are carried out for variable magnetic Prandtl number, ranging from $Pm=0.2$ up to $Pm=5$. \textcolor{black}{Four different values of the Ekman number are used in the present study: $Ek=(3 \times 10^{-6},10^{-5},3 \times 10^{-5},10^{-4})$. The Rayleigh number is varied up to approximately 50 times the critical Rayleigh number for the onset of non-magnetic convection.} We employ no-slip boundary conditions for the velocity field and electrically insulating boundary conditions for the magnetic field. The non-magnetic cases are identical to the dynamo cases except that $\mathbf{B}=0$.

%$\nu$ is the kinematic viscosity, $\Omega$ is the rotation rate of the system    %These are now both defined in the introduction, so I don't think they need to be defined again here.

We use the pseudo-spectral code Rayleigh to numerically solve the governing equations \citep{featherstone2022}. Rayleigh uses spherical harmonics to represent functions over shells, and Chebyshev polynomials to represent functions in radius. A full $3/2$ de-aliasing is used for both the spherical harmonics and the Chebyshev polynomials. Time stepping is performed with a 2nd-order Crank-Nicolson/Adams-Bashforth scheme. Rayleigh has been successfully benchmarked against the cases in \cite{uC01b}, and has been used in many previous studies \citep[e.g.][]{nF16,mC21,mH22,jN24}.

\subsection{Notation and outputs}
\label{S:notation}
In spherical geometries, large-scale, $\phi$-invariant structures can exist, such as zonal flows or the mean temperature profile. To separate these large-scale structures from the small-scale dynamics, we define the mean component of some field $X$ according to
\be
\overline{X} = \int_0^{2\pi} X \mathrm{d}\phi,
\ee
and the corresponding fluctuating component by $X'=X-\overline{X}$. The convective Reynolds number is computed as
\be
Re_c = \tavg{\sqrt{\langle \ub' \cdot \ub' \rangle}},
\ee
where $\langle \cdot \rangle$ is a volume average and $\tavg{\cdot}$ is a time average. Similarly, the Reynolds number calculated from the mean $\phi$-component of velocity, which we refer to as the zonal Reynolds number, is defined by
\be
\overline{Re}_\phi = \tavg{\sqrt{\langle \overline{\mathbf{u}}_\phi \cdot \overline{\mathbf{u}}_\phi \rangle}} .
\ee 

%I beleive we have removed all reference of magnetic spectra, so I am commenting this out
%We define $\mathcal{E}_l^m$ to be the kinetic energy density power spectrum at degree $l$ and order $m$, and $\mathcal{M}_l^m$ as the magnetic energy density power spectrum. Furthermore, we define the sum of these along $m$ to be
%\be
%E_l^{\prime} = \sum_{m=1}^l \mathcal{E}_l^m, \qquad M_l^{\prime} = \sum_{m=1}^l \mathcal{M}_l^m,
%\ee
%where the $m=0$ mode has been excluded in order to remove the influence of large-scale flows and fields.
We define $\mathcal{E}_l^m$ to be the kinetic energy density power spectrum at degree $l$ and order $m$ and $\mathcal{M}_l^m$ as the magnetic energy density power spectrum. Furthermore, we define the sum of these along $m$ to be
\be
E_l^{\prime} = \sum_{m=1}^l \mathcal{E}_l^m, \qquad M_l^{\prime} = \sum_{m=1}^l \mathcal{M}_l^m,
\ee
where the prime denotes that the $m=0$ mode has been excluded in order to remove the influence of large-scale flows and fields. $l_{peak}$ is defined as the degree at which $E^{\prime}_l$ achieves a maximum. To calculate $l_{peak}$, we use a simple polynomial interpolation of $E^{\prime}_l$ before finding the peak, as other studies have done \citep{cG19}. A peak length scale can then be defined as
\be
\ell^{\prime}_{peak} = \frac{\pi}{l_{peak}}.
\ee
%We no longer are using the spherical harmonic length scale from christensen, so I have commented it out.
%Similar to \cite{uC06}, we define the spherical harmonic length scales as
%\be
%\left(\ell_{sh}^v\right)^{-1} = \tavg{\left(\dfrac{\sum_{l=1}^{l_{max}} lE_l^{\prime}}{\pi\sum_{l=1}^{l_{max}} E_l^{\prime}}\right)}, \qquad \left(\ell_{sh}^b\right)^{-1} = \tavg{\left(\dfrac{\sum_{l=1}^{l_{max}} lM_l^{\prime}}{\pi\sum_{l=1}^{l_{max}} M_l^{\prime}}\right)}.
%\ee
%

The velocity and magnetic Taylor microscales are defined as
\be
\lambda_u^{\prime} = \tavg{\sqrt{\dfrac{\langle \ub'\cdot\ub' \rangle}{\langle\boldsymbol{\omega}' \cdot\boldsymbol{\omega}')\rangle}}}, \text{ and }  \lambda_b^{\prime} = \tavg{\sqrt{\dfrac{\langle \mathbf{B}'\cdot\mathbf{B}' \rangle}{\langle \mathbf{J}'\cdot\mathbf{J}'\rangle}}}, \label{E:tay}
\ee
respectively, where $\boldsymbol{\omega}=\nabla\times\mathbf{u}$ is the vorticity. The Taylor microscales are generally considered to be representative of the length scales at which diffusion (momentum/magnetic field) becomes important. For this reason, they are often referred to as dissipation length scales. Length scales are also computed directly from various terms in the momentum equation. We define the length scale computed from the advection term in the momentum equation as
\be
\ell_a^{\prime} =  \tavg{\langle(\mathbf{u}')^2\rangle/ \sqrt{\langle(\mathbf{u}'\cdot\nabla \mathbf{u}')_r^2\rangle}}, \label{E:adv}
\ee
and the length scale computed from the Lorentz force as
\be
\ell_l^{\prime} = \tavg{\langle (\mathbf{B}')^2 \rangle/ \sqrt{\langle(\mathbf{J}\hbox{\vphantom{J}}'\times\mathbf{B}\hbox{\vphantom{B}}')_r^2\rangle}}, \label{E:lor}
\ee
where $(\cdot)_r$ denotes the radial component of a vector. The viscous and Ohmic dissipation are calculated as
\be
\varepsilon_u = \tavg{\left(\nabla \times \ub\right)^2}, \qquad \varepsilon_b = \frac{1}{EkPm^2}\tavg{\left(\nabla \times \mathbf{B}\right)^2},
\ee
and the fraction of Ohmic dissipation is defined as 
\be
f_{ohm} = \frac{\varepsilon_b}{\varepsilon_b+\varepsilon_u}.
\ee
%The fraction of ohmic dissipation is then defined as
%\be
%f_{ohm} = \frac{\varepsilon_b}{\varepsilon_u+\varepsilon_b}.
%\ee
%I do not believe the total dissipatio or Nusselt number are shown in the paper anymore, so I have commented this out
%For the dynamo cases, the total dissipation is given by the sum of the viscous and ohmic dissipation: $\varepsilon_d=\varepsilon_u+\varepsilon_b$. For the convective cases, the total dissipation is simply equal to the viscous dissipation: $\varepsilon_c=\varepsilon_u$. The Nusselt number is calculated from the slope of the spherically averaged temperature at the outer boundary, as was done in previous studies \cite{tG16,jN24}. We denote the Nusselt number for dynamo cases as $Nu_d$ and the Nusselt number for convective cases as $Nu_c$. 

Since we are interested in understanding the small-scale (i.e.~convective-scale) force balance we remove the axisymmetric component of each force in the present study. Hereafter we refer to the resulting fluctuating forces simply as \textit{the} forces, unless otherwise specified.
For simplicity, only the radial component of the fluctuating forces are shown, with a primary focus on the volume integrated forces.
The rms of these force components are then computed according to
\be
\begin{aligned}
F_c^{\prime} &= \tavg{\sqrt{\langle \left(\frac{2}{Ek} \hz \times \ub'\right)_r^2\rangle}},  \quad F_p^{\prime} = \tavg{\sqrt{\langle \left(\frac{1}{Ek}\nabla P\hbox{\vphantom{P}}'\right)_r^2\rangle}},\\
F_b^{\prime} &= \tavg{\sqrt{\langle \left( \frac{Ra}{Pr} \lb \frac{r}{r_o} \rb T\hbox{\vphantom{T}}'\right)^2\rangle}},  \quad F_v^{\prime} = \tavg{\sqrt{\langle \left(\nabla^2 \ub'\right)_r^2\rangle}},\\
F_a^{\prime} &= \tavg{\sqrt{\langle \left(\ub \cdot \nabla \ub - \overline{\ub \cdot \nabla \ub}\right)_r^2\rangle}}, \quad F_t^{\prime} = \tavg{\sqrt{\langle \left(\partial_t \mathbf{u}'\right)_r^2\rangle}},\\
F_l^{\prime} &= \tavg{\sqrt{\langle \left(\frac{1}{EkPm} \left[\mathbf{J}\times \mathbf{B} - \overline{\mathbf{J}\times\mathbf{B}} \right]\right)_r^2\rangle}},\\
F_{ag}^{\prime} &= \tavg{\sqrt{\langle \left(Ek^{-1}\left[2\hat{z} \times \mathbf{u}' +\nabla P\hbox{\vphantom{P}}'\right]\right)_r^2\rangle}}.
\end{aligned}
\ee
We remove the $\ell=0$ part of the nonlinear terms since the $\ell=0$ component is not dynamically relevant in the radial component of the momentum equation for an incompressible fluid. Except for the buoyancy force which only appears in the radial direction and the viscous force which has significant boundary layers in the $\theta$ and $\phi$ directions, the different components of the forces exhibit identical asymptotic scalings, as can be deduced from the study of \cite{sN25}. As also shown in \cite{sN25}, the inclusion of Ekman boundary layers in the volume integration can have a significant influence on the observed balances. However, we find that the balances in the radial component of the momentum equation are not sensitive to whether the Ekman boundary layers are included, likely because the velocity normal to the boundary is asymptotically smaller than the horizontal velocity within the boundary layer \citep[e.g.][]{hG68}.

In the limit  $Ek \rightarrow 0$, it is well known that the critical Rayleigh number required to drive convection scales as $Ra_{c} = O(Ek^{-4/3})$ \citep{pR68}. Thus, the key control parameter in rapidly rotating convection-driven dynamos is the asymptotically reduced Rayleigh number defined as  \citep{kJ98a}
\be
\Rat = Ra Ek^{4/3} .
\ee 
Much of the analysis will be presented with this parameter.

Finally, we will use the notation $O(Ek^x)$ to denote that a quantity follows some particular asymptotic scaling with respect to the Ekman number, assuming that other parameters ($\Rat$, $Pr$, $Pm$) remain fixed.

\section{Quasi-geostrophic scalings}
\label{S:qg_scalings}

Here we briefly outline the known scalings from QG theory, which applies when the Coriolis force and the pressure gradient force enter at leading order, whereas all remaining terms in the momentum equation, including the Lorentz force, enter the dynamics at the next order \citep[e.g.][]{mC18}.
A simple derivation of the QG scalings in a plane layer can be derived by assuming that all of the terms in the fluctuating vorticity equation enter at the same asymptotic order in Ekman number. It is also assumed that length scales along the direction of the rotation axis are order one, and the length scales perpendicular to the rotation axis are all of the same asymptotic order.  \textcolor{black}{We note that multiple length scales are present in the radial direction in spherical geometries, though the larger length scale exhibits a weaker asymptotic dependence on the Ekman number \citep[e.g.][]{cJ00,eD04}; we discuss this point further in the results. Curling inertia and advection give respectively,
\be
\nabla \times \lb \dst \ubp \rb = O\left(\tau^{-1}  \ell^{-1} Re_c\right), \quad \nabla \times \lb \ubp \cdot \nabla \ubp \rb = O\left(Re_c^2 \ell^{-2}\right) ,
\ee
where $\tau$ is a characteristic timescale, $\ell$ is the lengthscale and $\ubp = O(Re_c)$. We have ignored the $\phi$-averaged component of the advection term since it is smaller than the term included above. For inertia and advection to be asymptotically comparable, we require the timescale to be
\be
\tau = O\lb \ell Re_c^{-1} \rb.
\ee
Curling the viscous and buoyancy forces gives, respectively,
\be
\nabla \times \lb \nabla^2 \ub \rb = O\left( \ell^{-3} Re_c \right) ,
\ee
\be
\nabla \times \lb \frac{Ra}{Pr} \frac{r}{r_o} T'  \, \hr \rb = O \lb \ell^{-1} Ra T' \rb,
\ee
where order one terms have been dropped.
%This implies
%\be
%O\left(Ek^{-1}Re\right) = O\left(Re^2 \ell^{-2}\right) = O\left(Re \ell^{-3}\right) = O\left(Ra T' \ell^{-1}\right),
%\ee
%where order one terms have been dropped and $\ell$ is the length scale. 
The above system of equations implies the following scalings: 
\be
Re = O \lb E^{-1/3} \rb, \quad T' = O \lb E^{1/3} \rb,
\ee
\be
\ell = O \lb E^{1/3} \rb, \quad \tau = O \lb E^{-2/3} \rb .
\ee
The above scalings are identical to those used in reduced models of rapidly rotating convection and rotating convection-driven dynamos \citep{mS06,mC15b}. Importantly, these scalings arise from the dominant (leading order) geostrophic force balance on the small $O(Ek^{1/3})$ horizontal length scale, and the theory is not limited to weakly nonlinear, or weakly forced, systems. The presence of $O(Ek^{1/3})$ in the above scalings does not arise from the assumption that the viscous force is dominant in the vorticity equation since the same dependence can arise without reference to viscosity. Indeed, $\Rat$ now controls the relative sizes of these terms; this point is addressed in the results.} %For a sphere, it is known that only the azimuthal direction has an $Ek^{1/3}$ length scale at the convective onset, while the cylindrical radial direction has an asymptotically larger length scale \citep{eD04}. 

In our non-dimensionalisation, the Coriolis force and pressure force are then expected to scale as $O(Ek^{-4/3})$, while the viscous force, buoyancy force, and advection term are expected to scale as $O(Ek^{-1})$. \textcolor{black}{These QG scalings are tested with the dynamo simulation data,} though it is not obvious a priori that the dynamo cases should follow these non-magnetic scalings \textcolor{black}{when the Lorentz force becomes comparable to the Coriolis force}. Further discussion of QG scalings in spherical convection can be found in \cite{jN24}.

%\FloatBarrier

\section{Results}
\label{S:results}

\subsection{Parameter regime and dependence on the magnetic Prandtl number}
\begin{figure*}
\begin{center}
%\subfloat[][]{\includegraphics[width=0.5\textwidth]{figures/Pm_forces_fluct_combined_v2_0.1_tick.eps}} %includes the 0.1 tick
%\subfloat[][]{\includegraphics[width=0.5\textwidth]{figures/Pm_dissipation_combined_0.1_tick.eps}}
\subfloat[][]{\includegraphics[width=0.5\textwidth]{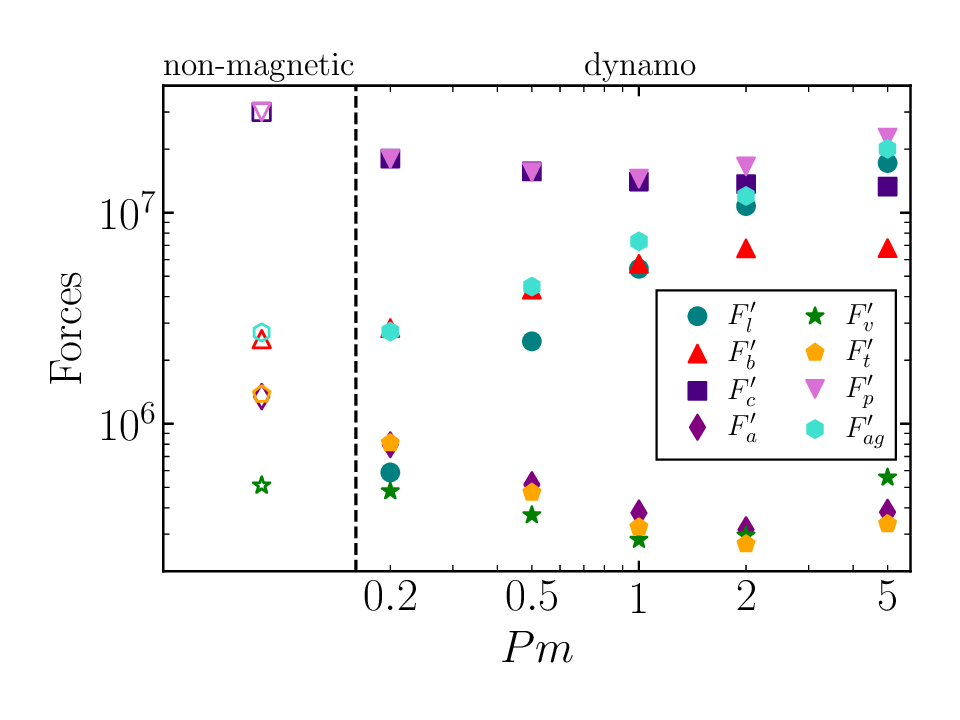}} %does not inlcude the 0.1 tick
\subfloat[][]{\includegraphics[width=0.5\textwidth]{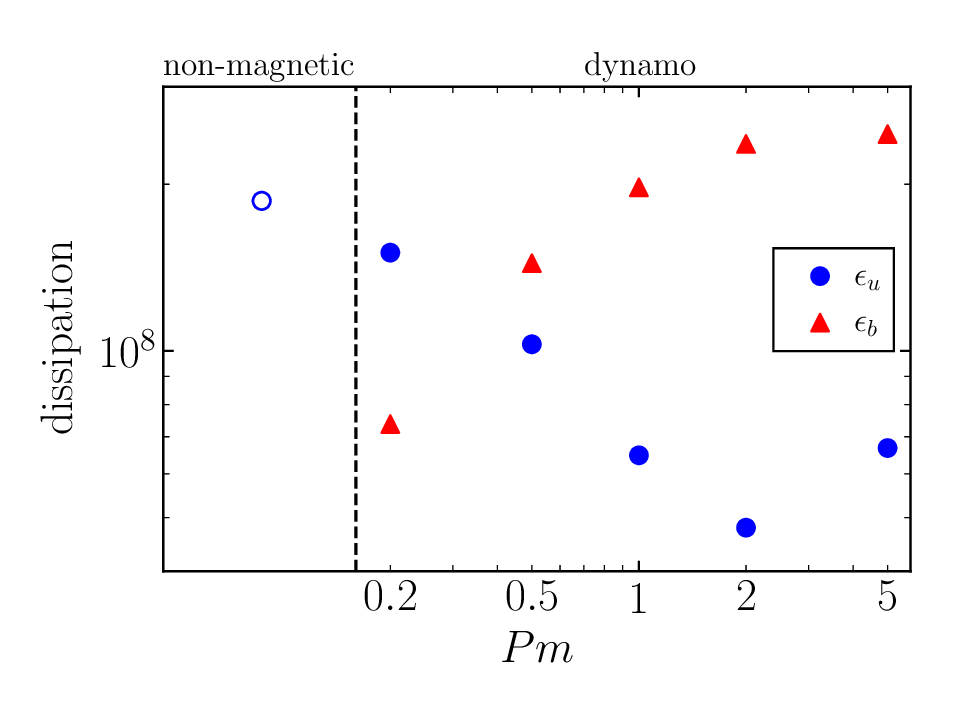}}
\caption{Influence of the magnetic Prandtl number ($Pm$) for fixed Ekman number ($Ek=10^{-5}$) and Rayleigh number ($Ra=1.5\times 10^{8}$) ($\Rat = Ra Ek^{4/3} =  32.3$): (a) radial components of the fluctuating forces; (b) viscous ($\epsilon_u$) and magnetic ($\epsilon_b$) dissipation. Points to the left of the vertical dashed lines are non-magnetic. }
\label{F:Pm_dependence}
\end{center}
\end{figure*}

As the force balance is known to vary with $Pm$, we start by showing the radial fluctuating force balance as a function of $Pm$ for the case $Ek=10^{-5}$ and $Ra=1.5\times 10^{8}$ in figure \ref{F:Pm_dependence}(a). \textcolor{black}{Note that when all other parameters are held fixed, the Lorentz force is controlled only by $Pm$.} The force balance for a non-magnetic case is shown to the left of the dashed line. At small values of $Pm$, the force balance for the dynamo cases is similar to the non-magnetic case. As expected, the Lorentz force increases with $Pm$ \citep{eD16,tS19,mM20}. At the largest value of $Pm=5$ reached here, the Lorentz force is slightly larger than the Coriolis force. The buoyancy force also increases with $Pm$, although less strongly than the Lorentz force -- the reason for this increase is discussed later. The remaining terms tend to decrease as $Pm$ is increased, which is likely due to a decrease in flow speeds as the dynamo becomes more efficient at converting kinetic energy to magnetic energy. It is therefore apparent that the dynamo cases have force balances ranging from QG to MAC depending on the value of $Pm$, which suggests that $Pm$ can affect the asymptotic order that the various forces enter in the momentum equation. In this paper, we will fix $Pm=2$, which leads to cases with a MAC force balance, especially at small Ekman numbers. Thus, these results can be contrasted with prior work on the asymptotics of small magnetic Prandtl number dynamos, which are essentially QG \citep{mY22}.

The viscous and ohmic dissipation are shown in figure \ref{F:Pm_dependence}(b) as a function of $Pm$. As $Pm$ is increased there tends to be a decrease in viscous dissipation and a corresponding increase in ohmic dissipation as the magnetic field becomes stronger.
% It is sometimes assumed that as $Pm$ becomes small, the ohmic dissipation dominates since the magnetic diffusivity becomes large compared to the kinematic viscosity. \textcolor{black}{Is there a reference for this?} However, our data suggests that the ohmic dissipation dominates at larger values of $Pm$, which is likely due to the magnetic field strength increasing with $Pm$. 
The range of values for the fraction of ohmic dissipation is $0.33 \le f_{ohm} \le 0.82$ for the cases shown in figure \ref{F:Pm_dependence}(b).

\subsection{Flow speeds}
\label{S:speed}

\begin{figure*}
\begin{center}
\subfloat[][]{\includegraphics[width=0.45\textwidth]{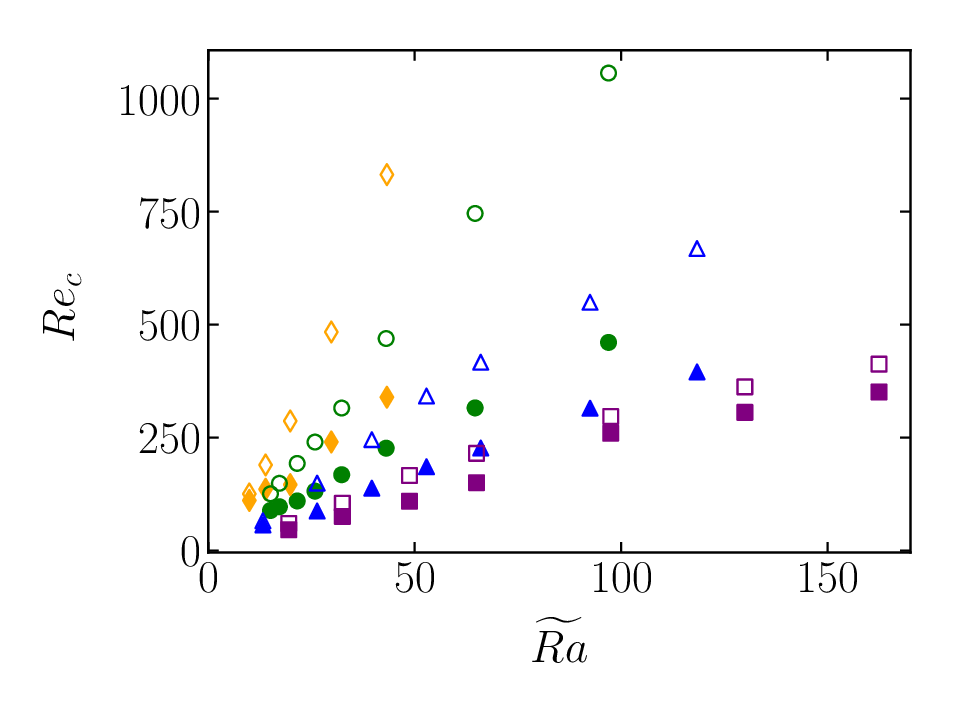}}
\subfloat[][]{\includegraphics[width=0.45\textwidth]{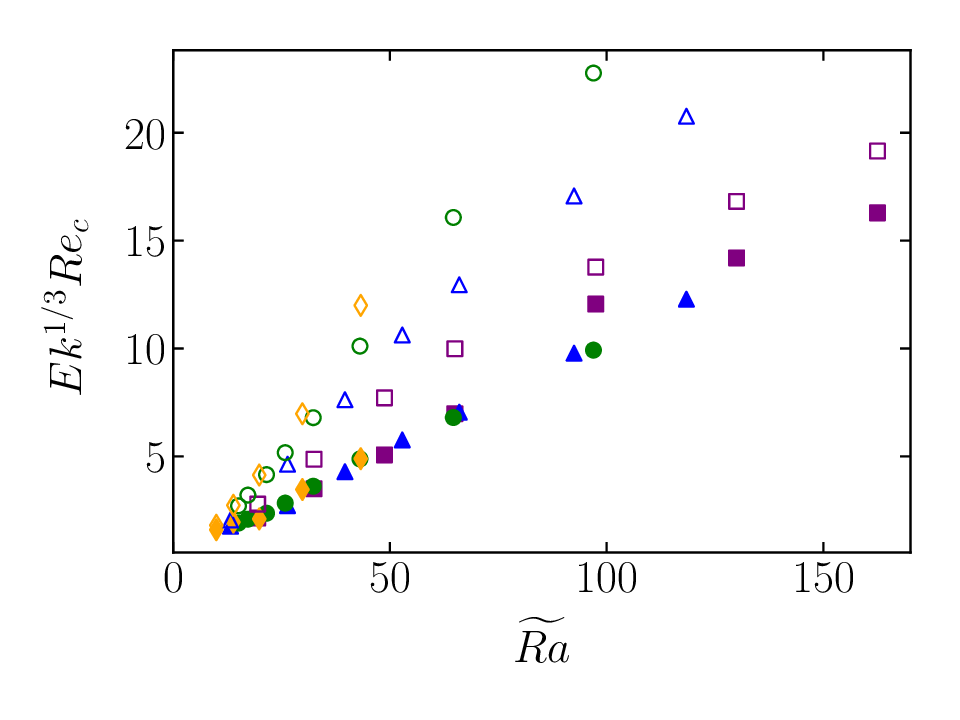}}\\
\subfloat[][]{\includegraphics[width=0.45\textwidth]{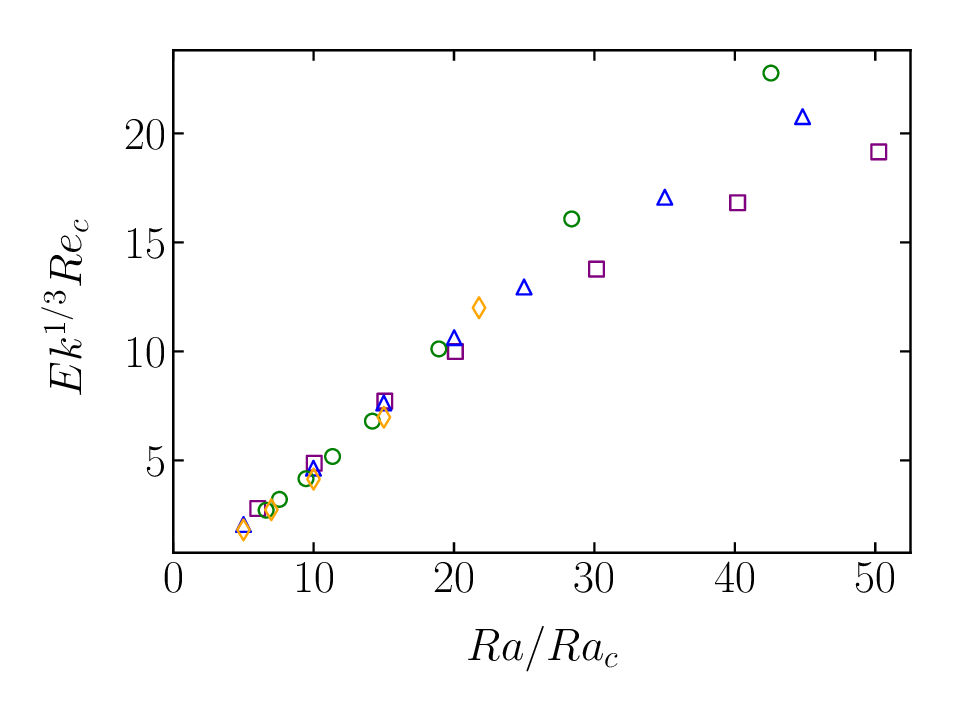}}
\subfloat[][]{\includegraphics[width=0.45\textwidth]{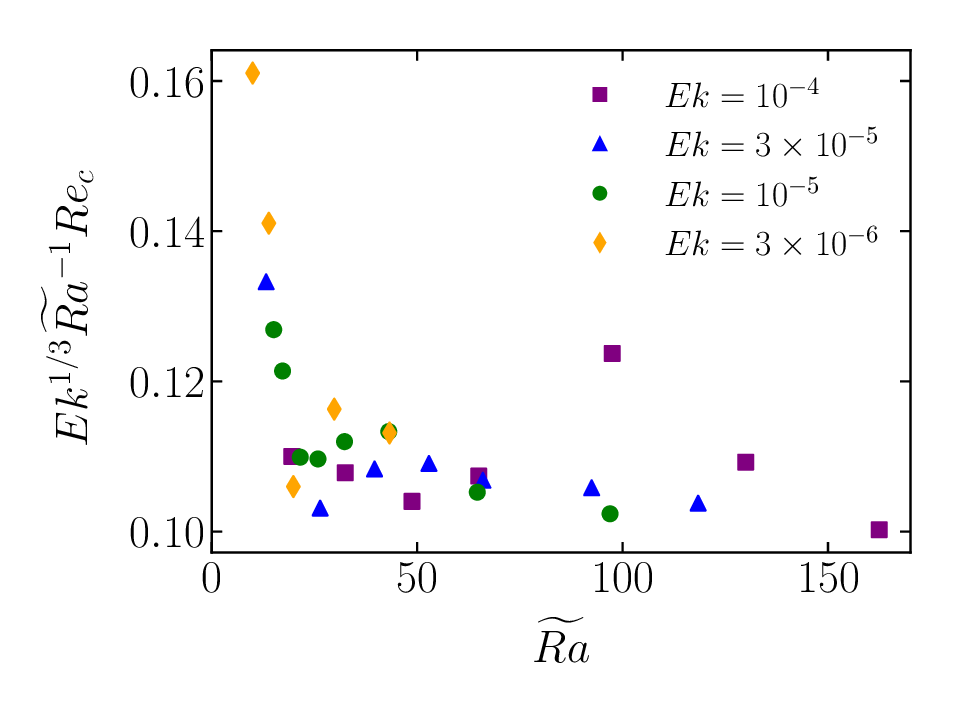}}\\
\subfloat[][]{\includegraphics[width=0.45\textwidth]{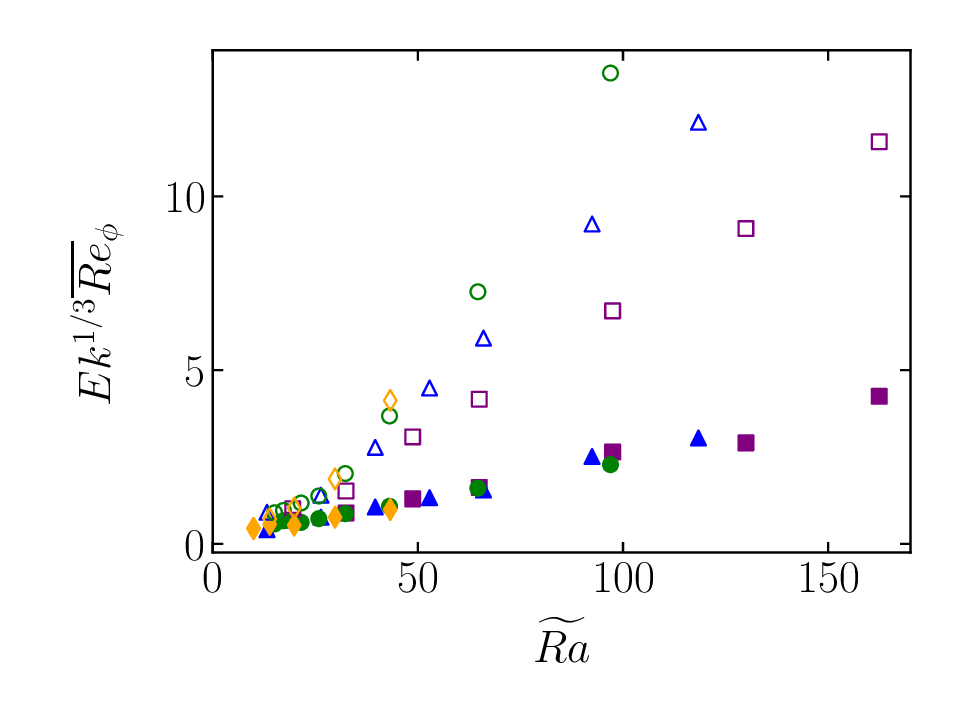}}
\subfloat[][]{\includegraphics[width=0.45\textwidth]{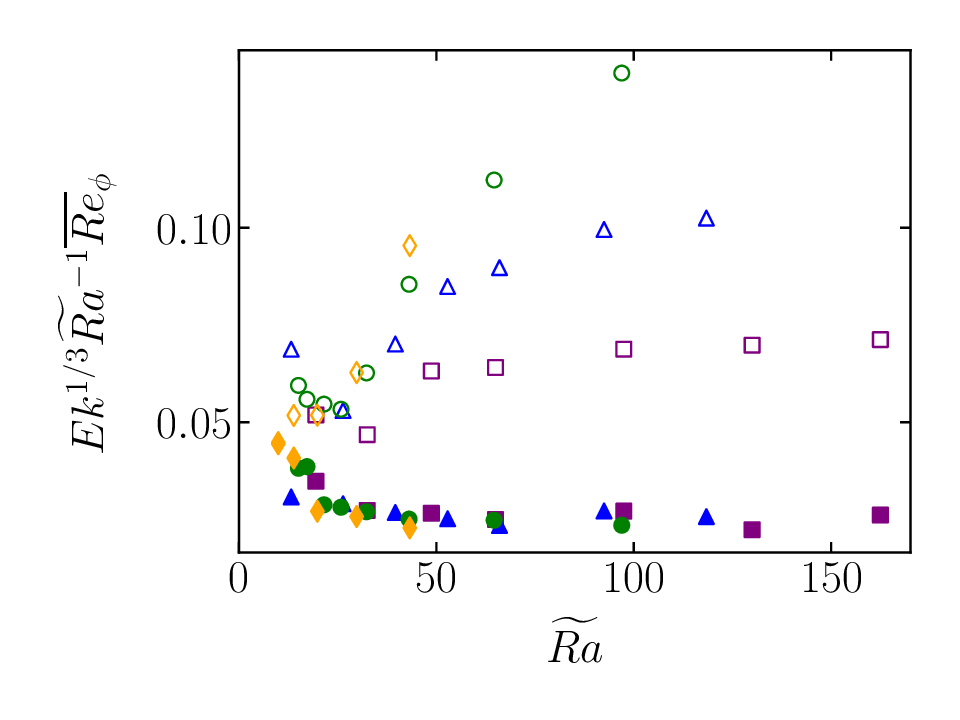}}
\caption{Reynolds number for both dynamo (filled symbols) and non-magnetic (open symbols) cases: (a) convective Reynolds number; (b) rescaled convective Reynolds number; (c) rescaled convective Reynolds number versus supercriticality, $Ra/Ra_c$; (d) compensated convective Reynolds number; (e) rescaled zonal Reynolds number; (f) compensated zonal Reynolds number.}
\label{F:Rec}
\end{center}
\end{figure*}

Figure \ref{F:Rec} shows the scaling of the Reynolds number for both dynamo (filled symbols) and non-magnetic (open symbols) simulations. The convective Reynolds number ($Re_c$) is shown in panels (a)-(d) and the zonal Reynolds number ($\overline{Re}_\phi$) is shown in panels (e) and (f). At small Rayleigh numbers, the convective Reynolds number for both the non-magnetic and dynamo cases are similar, although at higher Rayleigh numbers the non-magnetic cases are characterised by significantly larger $Re_c$. This difference can be understood in terms of the dynamo mechanism in which the magnetic field grows at the expense of the kinetic energy. We note that the difference in $Re_c$ between non-magnetic and dynamo cases becomes larger as $Ek \rightarrow 0$ and $\Rat$ is increased. 

\textcolor{black}{As indicated in figure \ref{F:Rec}(b), the dynamo cases are described well by the $O(Ek^{-1/3})$ scaling predicted from QG theory. The non-magnetic data appears to be less well described when viewed in these coordinates. To better understand this difference, we also plot the rescaled Reynolds number for the non-magnetic cases as a function of $Ra/Ra_c$ in figure \ref{F:Rec}(c), where good collapse is observed, indicating that the flow speeds are accurately described by the $O(Ek^{-1/3})$ scaling. Here, values of $Ra_c$ are taken from \cite{uC06}. We believe that the better collapse in this view of parameter space is because $Ra_c$ captures additional effects that may be important for the non-magnetic data, as opposed to using $\Rat$ alone. In particular, as described in \cite{eD04}, the critical Rayleigh number can be represented in terms of a composite asymptotic expansion, consisting of the well-known $Ek^{-4/3}$ leading order term, along with corrections of order $O(Ek^{-4/3} Ek^{2/9}) = O(Ek^{-10/9})$ and $O(Ek^{-4/3} Ek^{1/6}) = O(Ek^{-7/6})$ (see their equations 3.25), that arise from the effects of the long radial modulation scale and Ekman pumping, respectively. Why the dynamos do not seem to be influenced by these corrections remains unclear, though the magnetic data exhibits clear evidence for asymptotic scaling, even at relatively modest Ekman numbers.}

Noting the near linear trend of the convective Reynolds number for the dynamo cases in figure \ref{F:Rec}(b), we plot the convective Reynolds number compensated by $\Rat\vphantom{Ra}^{-1}$ in figure \ref{F:Rec}(d). The compensated data has much less dependence on $\Rat$, at least for sufficiently large values of $\Rat$. This scaling can be derived by balancing the Coriolis and buoyancy forces such that 
\be
- \frac{2\sin \theta}{Ek} u^{\prime}_{\phi} \sim \frac{Ra}{Pr} \lb \frac{r}{r_o} \rb T',   \quad \Rightarrow \quad u^{\prime}_{\phi} \sim Re_c \sim \frac{EkRa}{Pr} ,
\ee
if one assumes that the (fluctuating) temperature is order unity and independent of Rayleigh number. In terms of the reduced Rayleigh number this scaling becomes $Re_c \sim Ek^{-1/3} \Rat/Pr$. The data appears to be consistent with this scaling; further details of the force balance will be considered in the next section. 

The reduced zonal Reynolds number is shown in figure \ref{F:Rec}(e), where again the dynamo data appears better collapsed in comparison to the non-magnetic data. If we again assume a balance between the Coriolis and buoyancy forces, but now in the mean momentum equation, then we obtain an identical scaling to that given above for the convective Reynolds number, i.e.~$\overline{Re}_{\phi} \sim EkRa/Pr$. The compensated data given in figure \ref{F:Rec}(f) indicates that the dynamo cases are well described with this scaling, whereas the non-magnetic cases are not. \textcolor{black}{The mean force balance in dynamos is known to be well described by the thermal wind balance when the Lorentz force does not enter the leading order force balance \citep{jA05,mC21}. The mean force balance in the $\phi$-direction is between the Lorentz and Coriolis forces, though these components tend to be weaker in comparison to the leading order forces in the radial and colatitudinal directions \citep{jA05,mC21,rO21,yL25}.} However, for non-magnetic convection it is possible for the mean flows to be in geostrophic balance for sufficiently small Ekman number so that only large-scale viscous diffusion is available to saturate the zonal flow, as shown for non-magnetic convection using stress-free mechanical boundary conditions \citep{jN24}.

\subsection{Force scalings}
\label{S:forces}

\begin{figure*}
\begin{center}
\subfloat[][]{\includegraphics[width=0.5\textwidth]{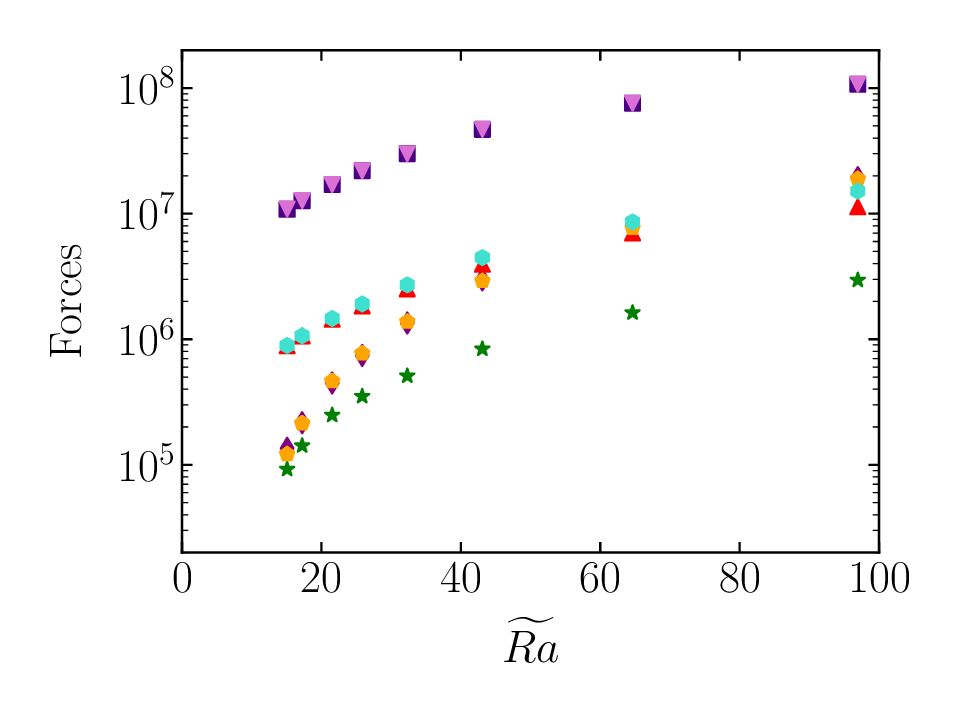}}
\subfloat[][]{\includegraphics[width=0.5\textwidth]{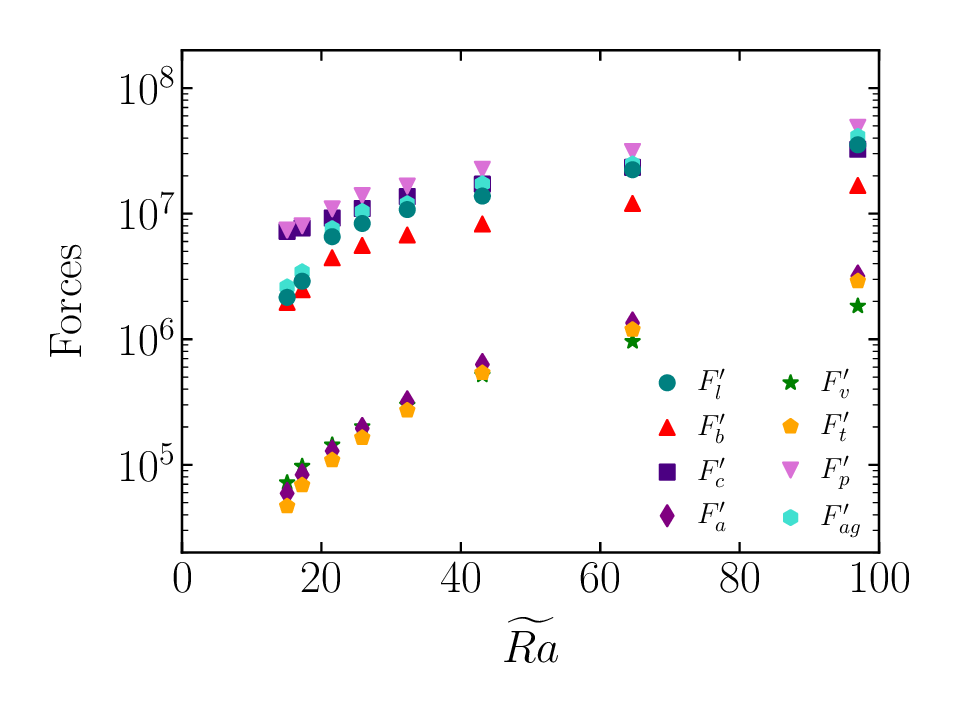}}
\caption{Forces versus $\Rat$ for (a) non-magnetic cases and (b) dynamo cases. The Ekman number is fixed at $Ek=10^{-5}$. }
\label{F:forces_Ek1e-5}
\end{center}
\end{figure*}

\begin{figure*}
\begin{center}
\subfloat[][]{\includegraphics[width=0.45\textwidth]{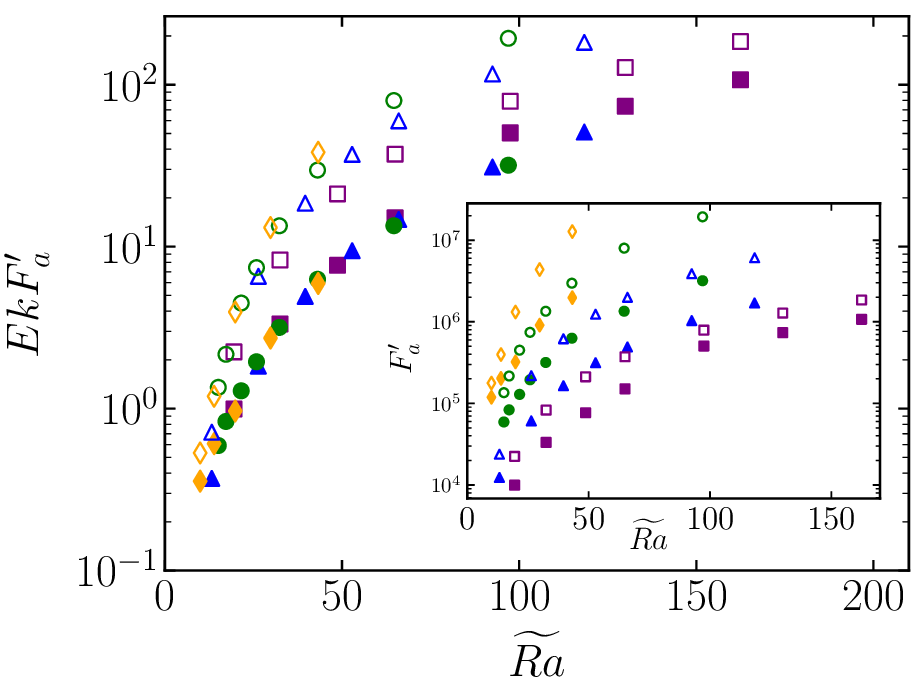}} \qquad
\subfloat[][]{\includegraphics[width=0.45\textwidth]{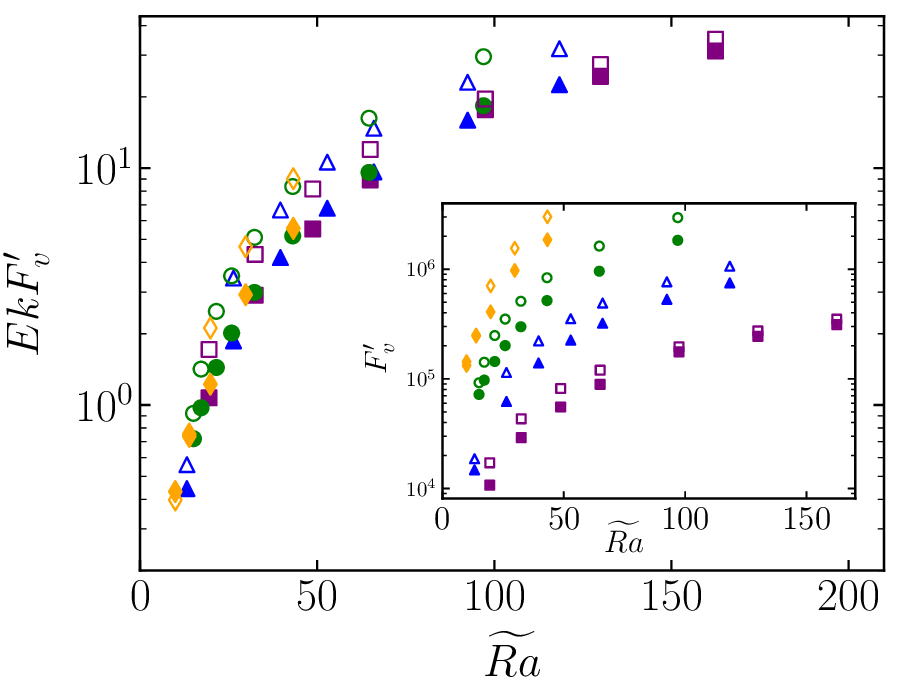}}\\
\caption{(a) Advection and (b) viscous force from the fluctuating radial momentum equation for all cases. The insets show the raw data with no rescaling. The symbols are the same as defined in figure \ref{F:Rec}.}
\label{F:forces}
\end{center}
\end{figure*}

\begin{figure*}
\begin{center}
\subfloat[][]{\includegraphics[width=0.45\textwidth]{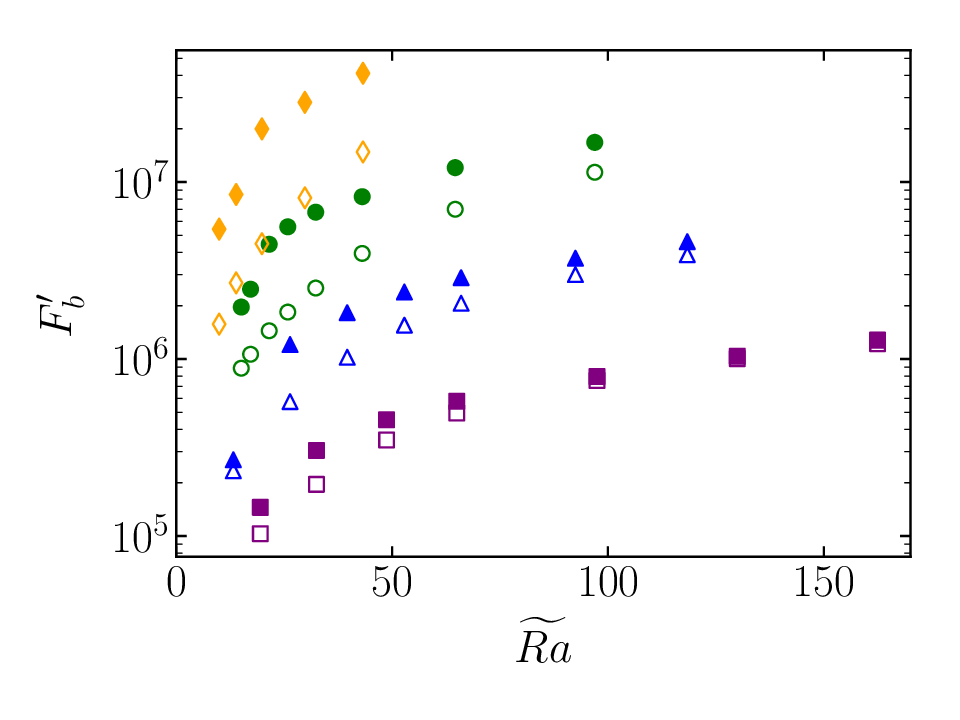}}
\subfloat[][]{\includegraphics[width=0.45\textwidth]{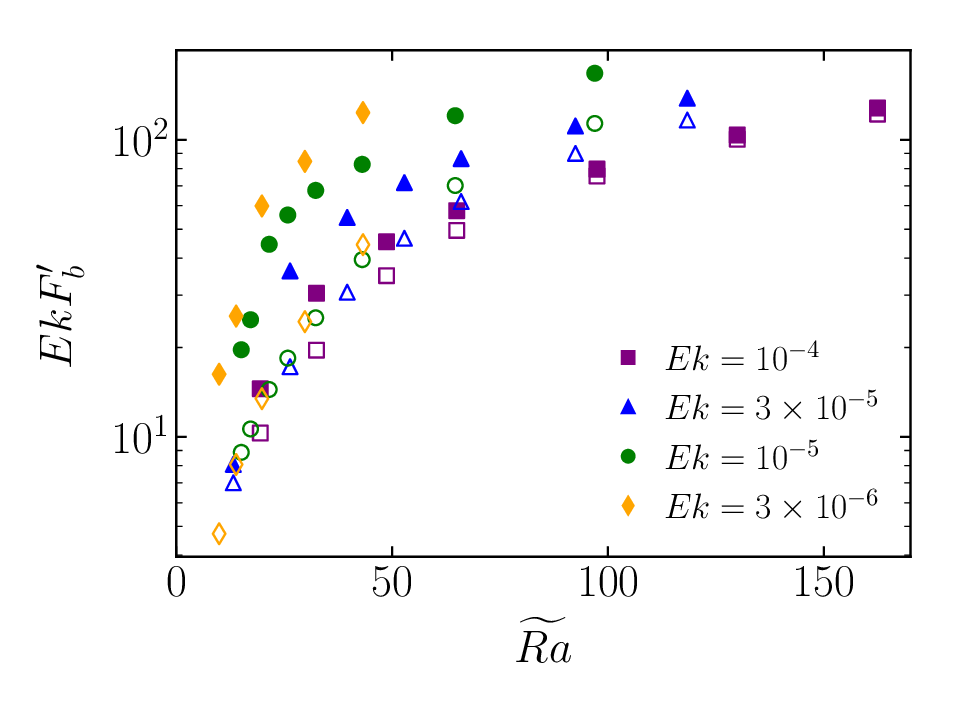}}\\
\subfloat[][]{\includegraphics[width=0.45\textwidth]{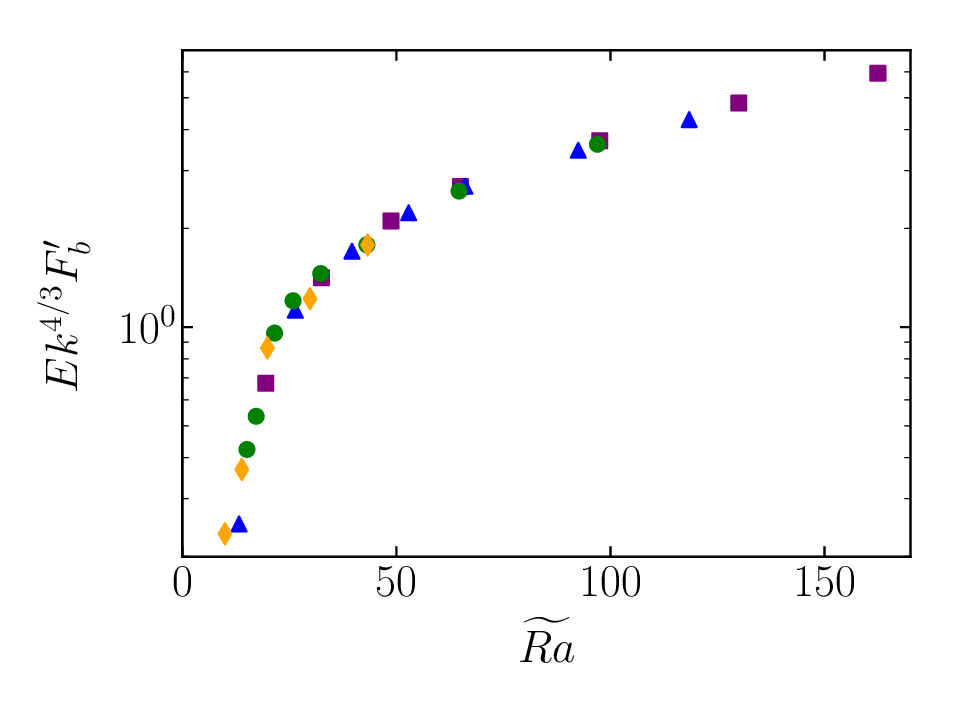}}
\subfloat[][]{\includegraphics[width=0.45\textwidth]{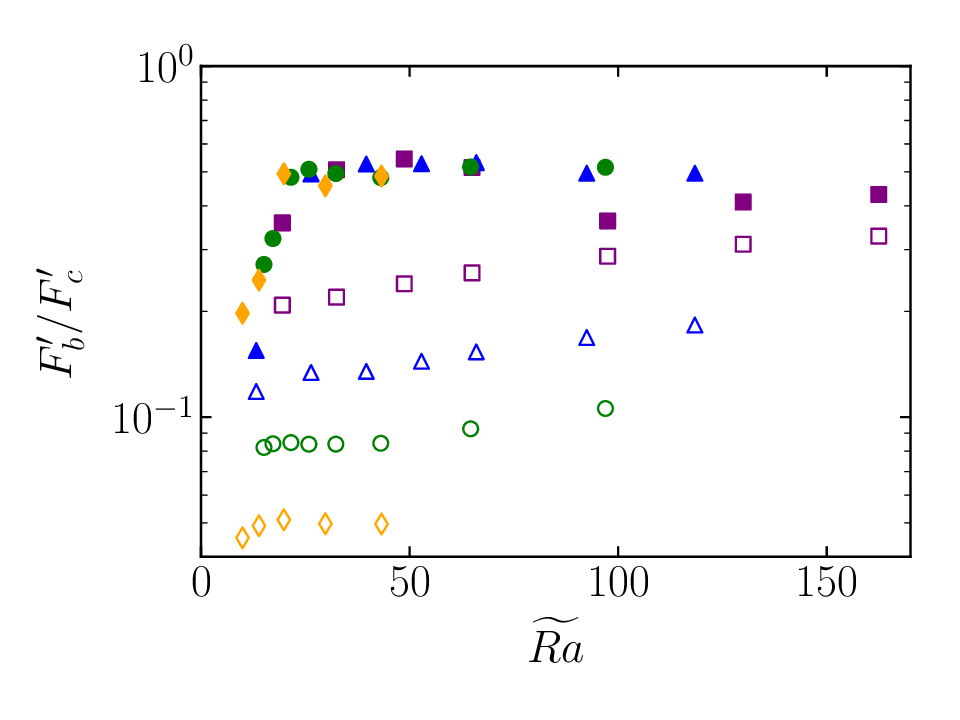}}
\caption{(a) Buoyancy force; (b) buoyancy force rescaled by $Ek$; (c) buoyancy force rescaled by $Ek^{4/3}$ for the dynamo cases; (d) ratio of the buoyancy force to the Coriolis force. The filled symbols denote dynamo cases and the unfilled symbols denote non-magnetic cases.}
\label{F:buoyancy}
\end{center}
\end{figure*}

In this section we compare the scalings of various terms from the radial component of the fluctuating momentum equation for the dynamo and non-magnetic cases. Figure \ref{F:forces_Ek1e-5} shows the forces as a function of $\Rat$ for $Ek=10^{-5}$. While the dominant balance for the non-magnetic cases is geostrophic, the dynamo cases have a dominant MAC balance between the Coriolis, pressure gradient, Lorentz, and buoyancy forces for $\widetilde{Ra} \gtrsim 20$. The dynamo cases have a larger fluctuating buoyancy force than the non-magnetic cases. In addition, the Coriolis force, viscous force, and advection term are all smaller for the dynamo cases than the non-magnetic cases, and this difference increases with $\Rat$. It is also interesting to note that the viscous force and advection term are similar in magnitude for the dynamo cases, but the advection term is larger than the viscous force and comparable to the buoyancy force for the high-$\Rat$ non-magnetic cases.

Figure \ref{F:forces} shows the advection term and the viscous force rescaled according to QG theory. Both the non-magnetic and dynamo cases are well collapsed by the same asymptotic scaling. This collapse is consistent with the scaling of the Reynolds number discussed in the previous section. Both terms are generally larger for the non-magnetic cases in comparison to the dynamo cases, which is likely due to the larger Reynolds numbers for the former.

The buoyancy force is plotted in figure \ref{F:buoyancy}(a) and the buoyancy force rescaled according to the QG prediction is shown in figure \ref{F:buoyancy}(b). While the non-magnetic cases are described well by this scaling, the dynamo cases show a systematic deviation away from this scaling as the Ekman number is reduced. In particular, the dynamo cases follow a stronger scaling closer to $O(Ek^{-4/3})$, as shown in figure \ref{F:buoyancy}(c). This scaling implies that the buoyancy force enters at the same asymptotic order as the Coriolis force for the dynamo cases. We test this scaling in figure \ref{F:buoyancy}(d), where we plot the ratio of the buoyancy force to the Coriolis force. For the non-magnetic cases, this ratio gets smaller as the Ekman number is decreased, as would be expected for a QG scaling. However, the dynamo cases reach a state where the buoyancy force is approximately half as large as the Coriolis force for all Ekman numbers presented here, indicating that the buoyancy force indeed follows the same scaling as the Coriolis force. 
%Note that, assuming a balance between the Coriolis and buoyancy forces implies $Re_c \sim (r/r_o)RaEk/Pr T'$. Dropping order one factors (and assuming the fluctuating temperature is order one) gives $Re_c \sim RaEk/Pr = Ek^{-1/3}\Rat/Pr$. This is possibly the reason for the $O(Ek^{-1/3})$ scaling of the Reynolds number even though the force balance for the dynamo simulations is not QG. 
As previously mentioned, this balance in the dynamos leads to a Reynolds number scaling that is diffusion-free and agrees with the scaling shown in figure \ref{F:Rec}(d). This behaviour is very different from the non-magnetic cases, where the buoyancy force is asymptotically smaller than the Coriolis force. The cause for this difference in scaling between the dynamo and non-magnetic cases can be explained from a balance between the power generated through buoyancy and dissipation, which will be explored further in the next section.

\begin{figure*}
\begin{center}
\subfloat[][]{\includegraphics[width=0.45\textwidth]{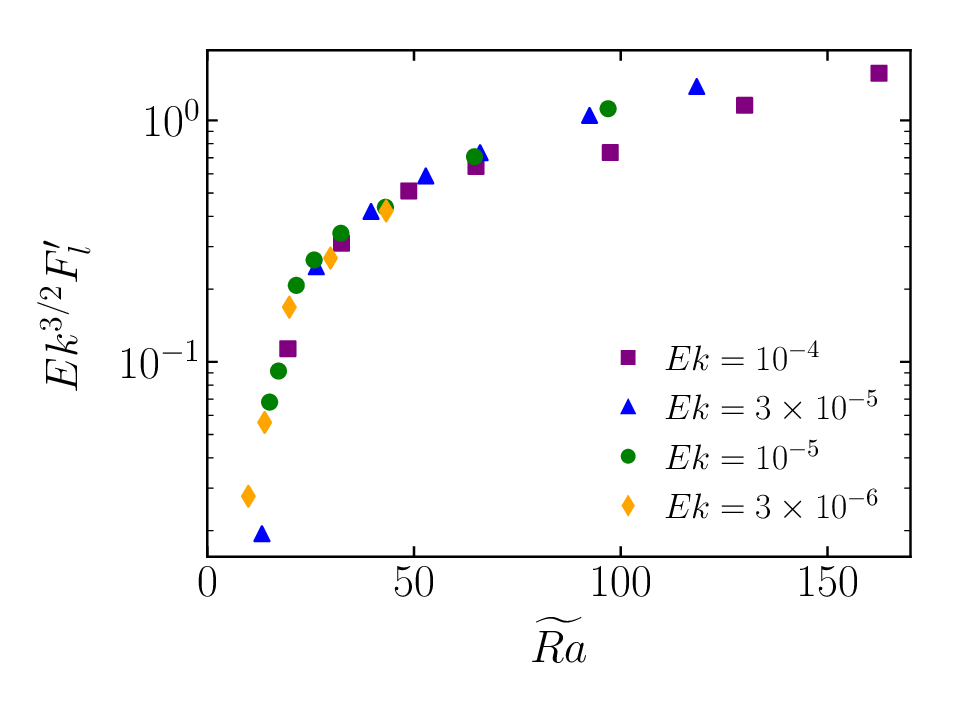}}
\subfloat[][]{\includegraphics[width=0.45\textwidth]{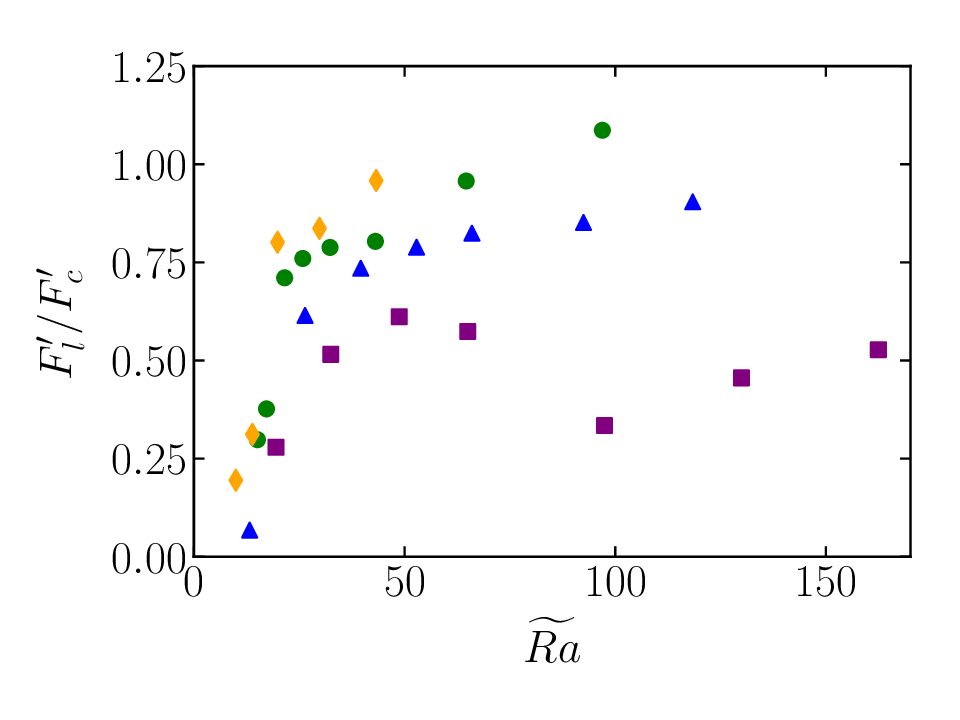}} \\
\subfloat[][]{\includegraphics[width=0.45\textwidth]{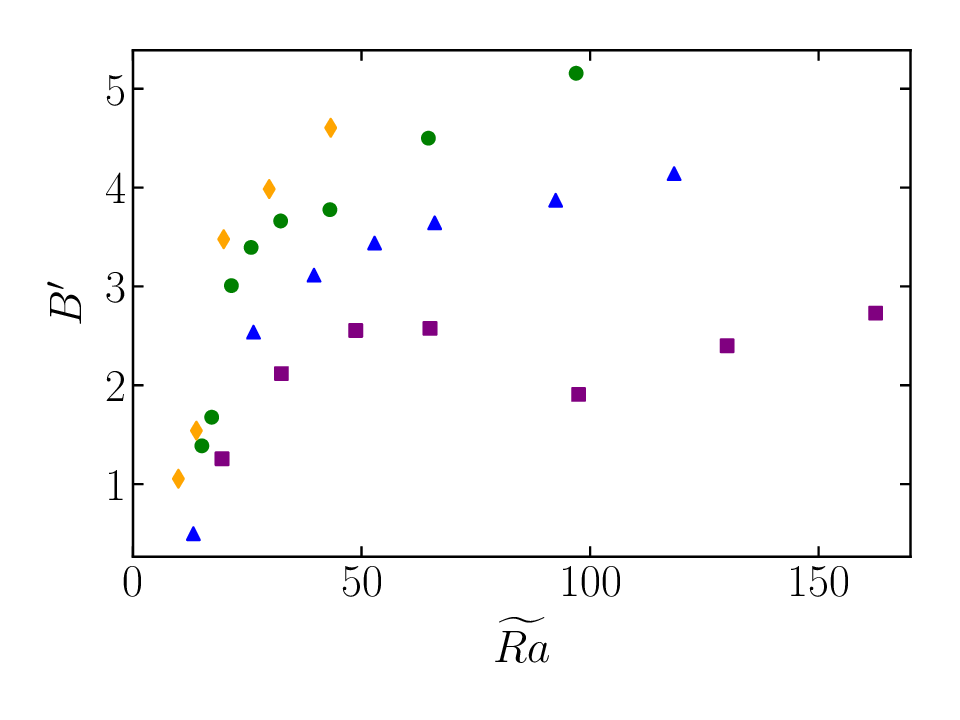}}
\subfloat[][]{\includegraphics[width=0.45\textwidth]{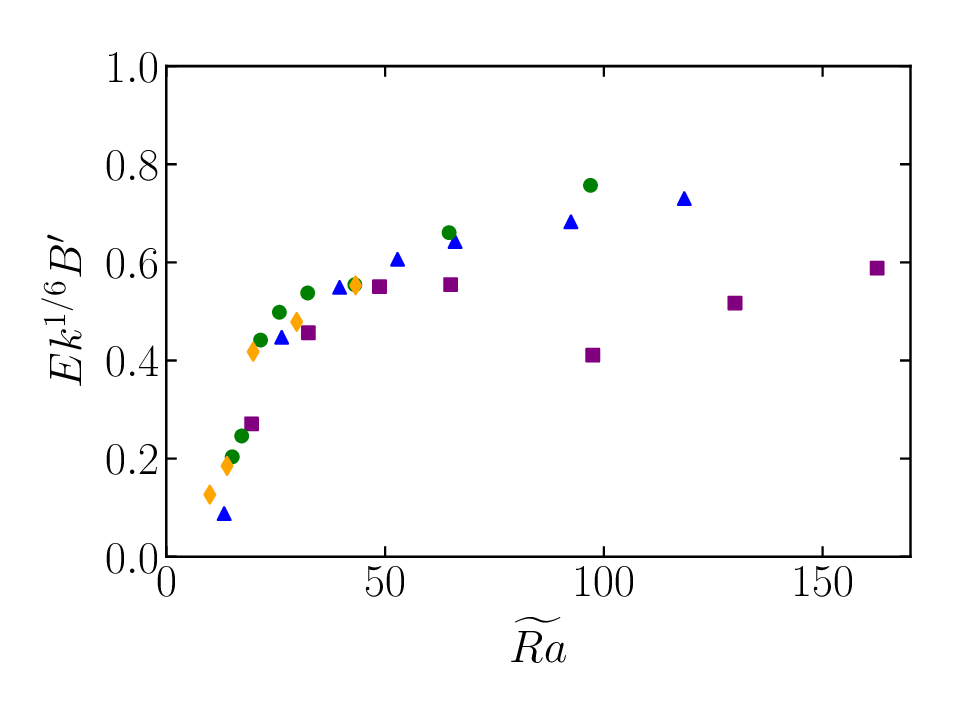}}
\caption{(a) Rescaled Lorentz force; (b) ratio of the Lorentz force to the Coriolis force; (c) rms fluctuating magnetic field; (d) rescaled fluctuating magnetic field.}
\label{F:lorentz}
\end{center}
\end{figure*}

The rescaled Lorentz force is shown in figure \ref{F:lorentz}(a), where an empirical scaling of $O(Ek^{-3/2})$ has been used to collapse the data. The ratio of the Lorentz force to the Coriolis force is shown in figure \ref{F:lorentz}(b), where it can be seen that this ratio increases with decreasing $Ek$. However, note that while the Lorentz force is scaling more strongly with Ekman number than the Coriolis force, the Coriolis force is larger than the Lorentz force for most of the cases. %This observation suggests that as the Ekman number becomes smaller there is either a change in scaling, or that the Lorentz force would become larger than the Coriolis force. 
It is important to recall that $Pm$ is constant in these cases; as shown in figure \ref{F:Pm_dependence}, the strength of the Lorentz force depends strongly on $Pm$. We also note that this scaling of the Lorentz force together with the scaling of the length scale of the Lorentz force of $O(Ek^{1/6})$ (shown later) implies that the magnetic field scales as $O(Ek^{-1/6})$, which is different than the order one scaling proposed in \cite{mC21}. A plot of the rms of the fluctuating magnetic field is provided in figure \ref{F:lorentz}(c), and the corresponding rescaled data is given in figure \ref{F:lorentz}(d).

\begin{figure*}
\begin{center}
\subfloat[][]{\includegraphics[width=0.45\textwidth]{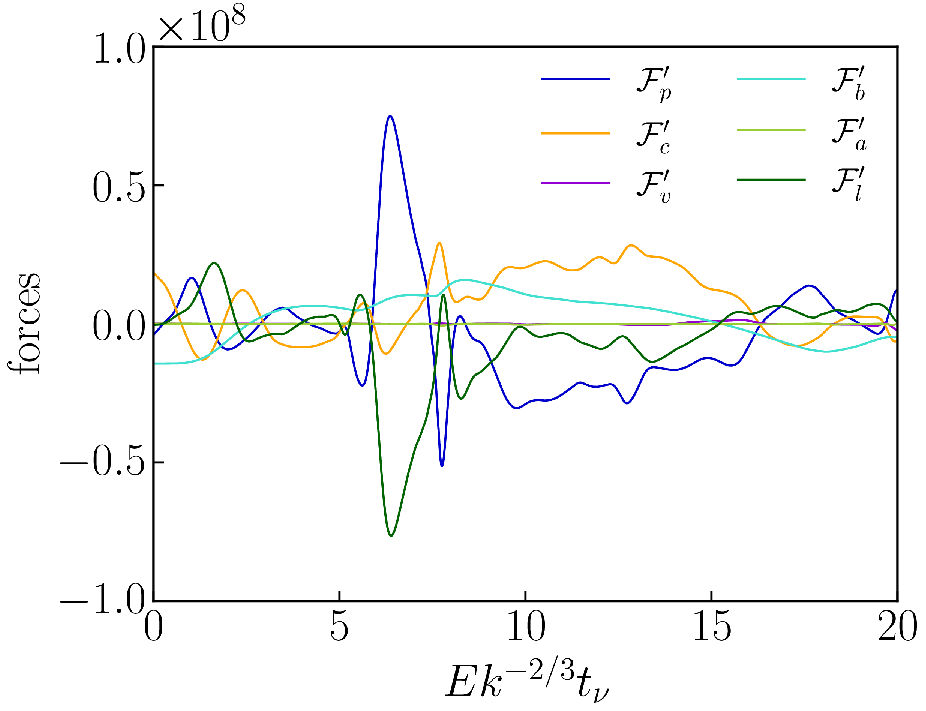}} \quad
\subfloat[][]{\includegraphics[width=0.45\textwidth]{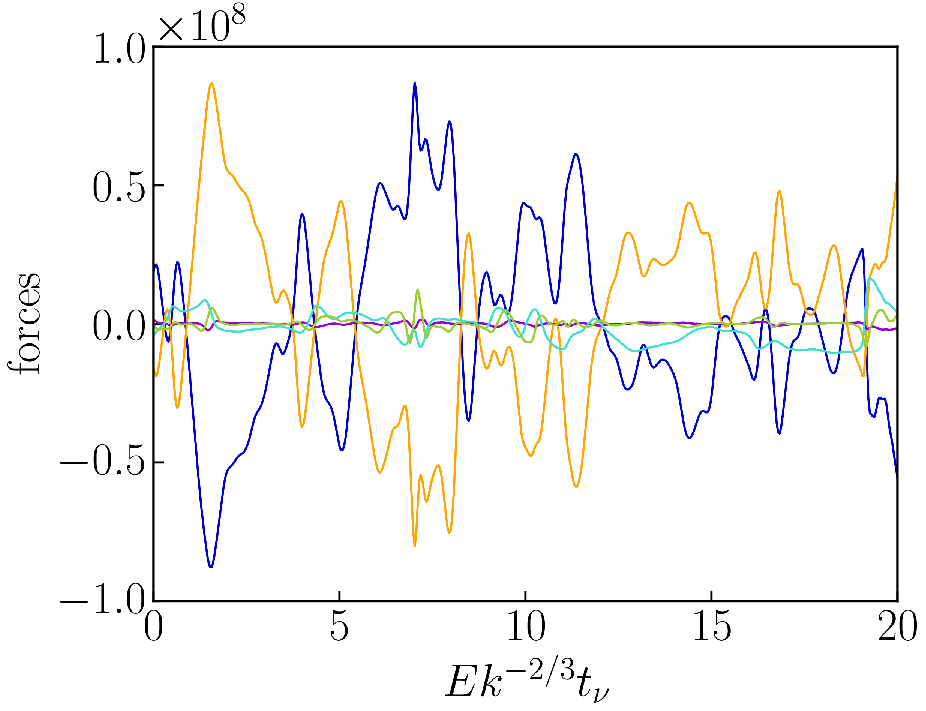}}\\
\caption{Temporal behaviour of the forces in the equatorial plane near the radial midpoint for $Ek=10^{-5}$ and $Ra=1.5\times 10^{8}$ ($\Rat = 32.3$): (a) dynamo; (b) non-magnetic. Time is in units of small-scale viscous diffusion time $Ek^{-2/3}t_\nu$ where $t_\nu$ is the large-scale viscous diffusion time. The labels $\mathcal{F}_p^{\prime}$, $\mathcal{F}_c^{\prime}$, $\mathcal{F}_v^{\prime}$, $\mathcal{F}_b^{\prime}$, $\mathcal{F}_a^{\prime}$, and $\mathcal{F}_l^{\prime}$ denote the fluctuating radial pressure gradient force, Coriolis force, viscous force, buoyancy force, advection, and Lorentz force, respectively.} 
\label{F:point_probe}
\end{center}
\end{figure*}

\begin{figure*}
\begin{center}
\subfloat[][]{\includegraphics[width=0.5\textwidth]{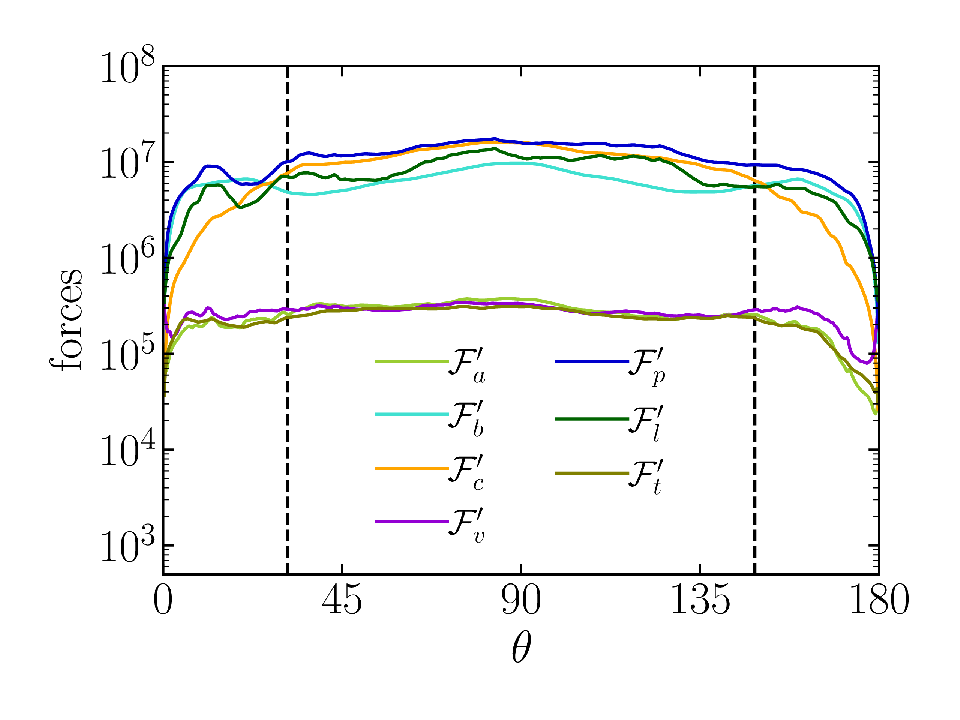}}
\subfloat[][]{\includegraphics[width=0.5\textwidth]{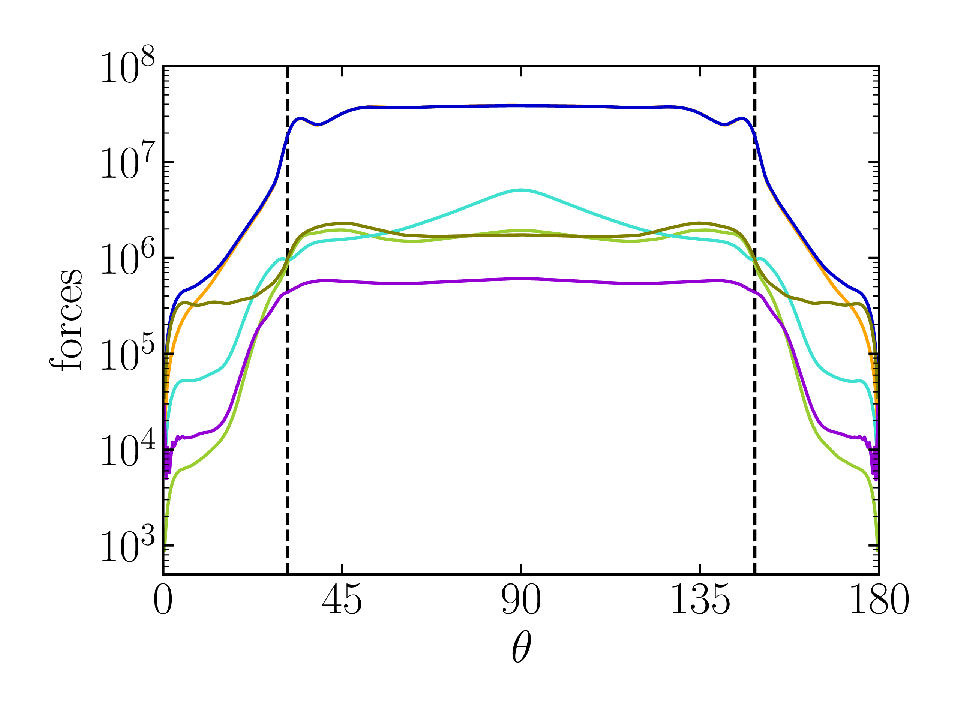}}\\
\caption{Force balance near the midpoint between the shells as a function of latitude for $Ek=10^{-5}$ and $Ra=1.5\times 10^{8}$ ($\Rat = 32.3$). Both a time-average and a rms in longitude ($\phi$) have been performed. The dashed vertical lines denote the location of the tangent cylinder. The cases shown are (a) the dynamo case and (b) the non-magnetic case. }
\label{F:meridional_forces}
\end{center}
\end{figure*}

The temporal behaviour of the fluctuating forces is shown in figure \ref{F:point_probe} for both a non-magnetic and dynamo case at $Ek=10^{-5}$, $Ra=1.5\times 10^{8}$ ($\Rat = 32.3$) at a fixed point in the equatorial plane that lies midway between the shells. In our non-dimensionalisation, time is in units of the large-scale viscous diffusion time, $t_\nu$. However, we choose to plot the data in terms of the small-scale viscous diffusion time $Ek^{-2/3}t_\nu$, which is the dynamically relevant timescale for QG convection \citep[e.g.][]{tO23,tO25}. For the non-magnetic case, the leading-order force balance is between the Coriolis force and the pressure gradient force, as would be expected for QG convection. The dynamo case is well described by a MAC balance in which the Coriolis force, pressure force, buoyancy force, and Lorentz force are all comparable and leading-order in the dynamics. Although these four terms tend to be the largest in a time averaged sense, the force balance exhibits significant changes with time. For example, at time $Ek^{-2/3}t_\nu \approx 6$, there is an approximate balance between the Lorentz and pressure forces, while at other times the Coriolis force is larger than the Lorentz force. Therefore, the leading-order force balance for the non-magnetic cases and the dynamo cases is different, as suggested by the asymptotic scalings for these forces. We note that other studies find the force balance might depend on length scale \citep[e.g.][]{tS19,jA17,tT21}, with many studies finding that the QG balance holds only at large length scales for dynamo simulations. However, the data shown here indicates that the pointwise balance is not geostrophic.

Further information on the spatial variation of these forces is provided in figure \ref{F:meridional_forces}, which shows the radial fluctuating forces as a function of latitude for the same cases as in figure \ref{F:point_probe}. For the non-magnetic case, a dominant balance between the Coriolis force and pressure gradient force exists across almost all latitudes. The advection term is larger than the viscous term outside the tangent cylinder, whereas the opposite is true  inside the tangent cylinder. This difference is likely due to the higher Rayleigh numbers needed for convection within the tangent cylinder \citep[e.g.][]{tG23}. For the dynamo case, a dominant balance between the Coriolis force, pressure gradient force, Lorentz force, and buoyancy force prevails across all latitudes, except near the poles where the Coriolis force becomes small. Therefore, for both the dynamo and non-magnetic cases, the dominant force balance does not vary strongly with latitude, although the absolute sizes of the various terms are sensitive to latitude. Most notably, there is a decrease in the size of the Coriolis, pressure gradient, viscous, and buoyancy forces by a few orders of magnitude inside the tangent cylinder for the non-magnetic case. 

One of the key points elucidated by this data is that the advection term is comparable to both inertia term and the viscous force throughout the volume for both the dynamo and convection cases. If we use the asymptotic scalings for the velocity and length scale as $Ek^{-1/3}$ and $Ek^{1/3}$, respectively, then advection and the viscous force both scale according to (as confirmed in figure \ref{F:forces})
\be
\ubp \cdot \nabla \ubp = O(Ek^{-1}), \quad \text{and} \quad \nabla^2 \ubp = O(Ek^{-1}).
\ee
\textcolor{black}{As done in section 3, if we define $\tau$ as the relevant dynamical timescale and balance inertia with either of these terms then we have
\be
\partial_t \ubp \sim u \tau^{-1} = O(Ek^{-1}) \quad \Rightarrow \quad  \tau = O \lb Ek^{2/3} \rb .
\ee
Interestingly, this is the asymptotic scaling for the period of convective Rossby waves, implying that these dynamics persist despite the leading order MAC balance.}

\subsection{Temperature and dissipation relations}

\begin{figure}
\begin{center}
\subfloat[][]{\includegraphics[width=0.50\textwidth]{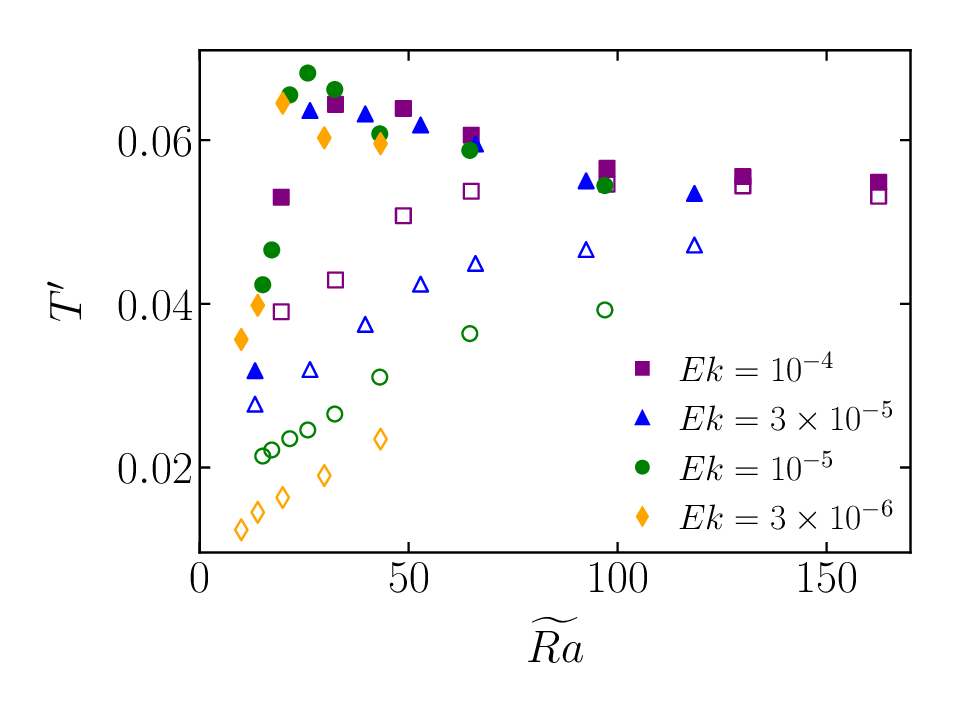}}
\caption{rms fluctuating temperature. The filled symbols denote dynamo cases and the open symbols denote non-magnetic cases.}
\label{F:temp_rms}
\end{center}
\end{figure}

Analysis of the forces in section \ref{S:forces} shows that the fluctuating buoyancy force, and therefore the fluctuating temperature, is characterised by a unique asymptotic scaling depending on whether the magnetic field is present. Figure \ref{F:temp_rms} shows the rms of the fluctuating temperature for all cases, where this difference can be observed -- for non-magnetic rotating convection the magnitude of the fluctuating temperature scales according to $Ek^{1/3}$, whereas the dynamo cases exhibit a fluctuating temperature that becomes independent of both $Ek$ and $\Rat$, as has been noted previously \citep{mC21}. This difference can be explained based on energetic arguments as follows. For the dynamo cases, a dissipation equation can be derived by dotting the full momentum equation with the velocity field and averaging over both the fluid volume ($V$) and time to give
\be
\begin{aligned}
\int_V &u_rf_b\, dV = \int_V u'_rf'_b+\overline{u}_r\overline{f}_b \, dV = \int_V \boldsymbol{\omega}^2 \, dV \\&+\frac{1}{EkPm^2} \int_V \mathbf{J}^2 \, dV ,
\label{E:buoyancy_power}
\end{aligned}
\ee
where $f_b=(Ra/Pr)(r/r_o)T$ is the buoyancy force.
%$S$ consists of the two spherical surfaces enclosing the domain, and $\hat{\mathbf{n}}$ is $\hat{\mathbf{r}}$ on the outer surface and $-\hat{\mathbf{r}}$ on the inner surface. Note that there is an implicit time average in this equation. 
%From left to right, the three integrals represent the power supplied by buoyancy, the viscous dissipation, and the Ohmic dissipation, respectively.
From left to right, the first two integrals represent the power supplied by buoyancy, the third integral represents the viscous dissipation, and the fourth integral represents the Ohmic dissipation.
% (the Poynting flux at the boundaries. The time-average Poynting flux at the boundaries vanishes if the total magnetic energy outside the fluid domain is bounded, so the Poynting flux should be irrelevant on time average. 
In the non-magnetic cases, the same equation can be derived, but without the Ohmic dissipation term. For our cases we find that $\langle \overline{u}_r\overline{f}_b\rangle \ll \langle u'_r f'_b \rangle$ such that an approximate balance in the non-magnetic cases is present between the fluctuating buoyancy power generation and viscous dissipation,
%where $\overline{f}_b=(Ra/Pr)(r/r_o)\overline{T}$ is the mean buoyancy force.
\be
\langle u'_r f'_b \rangle \sim \langle \boldsymbol{\omega}^2 \rangle = \langle \mathbf{u} \cdot \mathbf{f}_v\rangle,
\ee 
where $\mathbf{f}_v=\nabla^2 \mathbf{u}$, and we recall that $\langle \rangle$ denotes a volume average. Assuming that viscous dissipation does not become entirely concentrated in the boundary layer, which we confirmed for some cases, it would seem likely that the fluctuating buoyancy force cannot follow a stronger scaling than the viscous force. This is what was observed in subsection \ref{S:forces}, where both the fluctuating buoyancy force and the fluctuating viscous force follow the same Ekman number scaling for the non-magnetic cases. Using $\mathbf{f}_v = O(Ek^{-1})$, the balance between the viscous force and fluctuating buoyancy force implies $(Ra/Pr)(r/r_o)T' = O(Ek^{-1})$. Dropping order one terms and using $Ra=O(Ek^{-4/3})$, the fluctuating temperature must scale as $T'=O(Ek^{1/3})$.

The situation is very different for the dynamo simulations where the Ohmic dissipation exceeds the viscous dissipation. As a result, the fluctuating buoyancy force can be much larger than the viscous force such that the fluctuating temperature can be asymptotically larger than $O(Ek^{1/3})$. This change in the scaling of the fluctuating temperature is observed in figure \ref{F:temp_rms}, where the fluctuating temperature for the dynamo cases is order one in the sense that it does not depend on the Ekman number. This is the strongest scaling the fluctuating temperature can follow, since the fluctuating temperature is always between the temperature on the outer and inner boundary, which is between zero and one here.

Since the fluctuating temperature for the dynamo cases is independent of the Ekman number, and the fluctuating velocity scales as $O(Ek^{-1/3})$, the power generated through buoyancy would be expected to scale as 
%\be
%P_d \sim \langle u'_r F'_b\rangle \sim O(Ek^{-1/3})O(Ek^{-4/3}) \sim O(Ek^{-5/3}).
%\ee
\be
P_d \sim \langle u'_r F'_b\rangle \sim O(Ek^{-5/3}).
\ee
Unless the length scale is much smaller for the dynamo cases than the non-magnetic cases (which we do not observe), this power generated through buoyancy can only be balanced by the ohmic dissipation, which leads to a predicted ohmic dissipation scaling of $O(Ek^{-5/3})$. For the non-magnetic cases, the flow speeds scale as $Re_c \sim O(Ek^{-1/3})$ and the dissipation length scale scales as $O(Ek^{1/3})$, so the viscous dissipation for the non-magnetic cases should scale as $O(Ek^{-4/3})$. This line of reasoning therefore suggests that the ohmic dissipation is asymptotically larger than the viscous dissipation when the MAC balance is achieved. 
Though not shown, these scalings for both the viscous and ohmic dissipation were confirmed with the simulation data. 
These considerations imply that the dynamo simulations have an asymptotically larger dissipation than the non-magnetic cases, although in the parameter range we have explored, the total dissipation for both the dynamo and non-magnetic cases are similar. At larger values of the Ekman number the total dissipation is approximately the same for both the dynamo and non-magnetic cases. As $Ek$ is reduced the dynamo simulations have greater total dissipation than the non-magnetic cases by a factor up to approximately 2.5.

\subsection{Length scales}

\begin{figure*}
\begin{center}
\subfloat[][]{\includegraphics[width=0.45\textwidth]{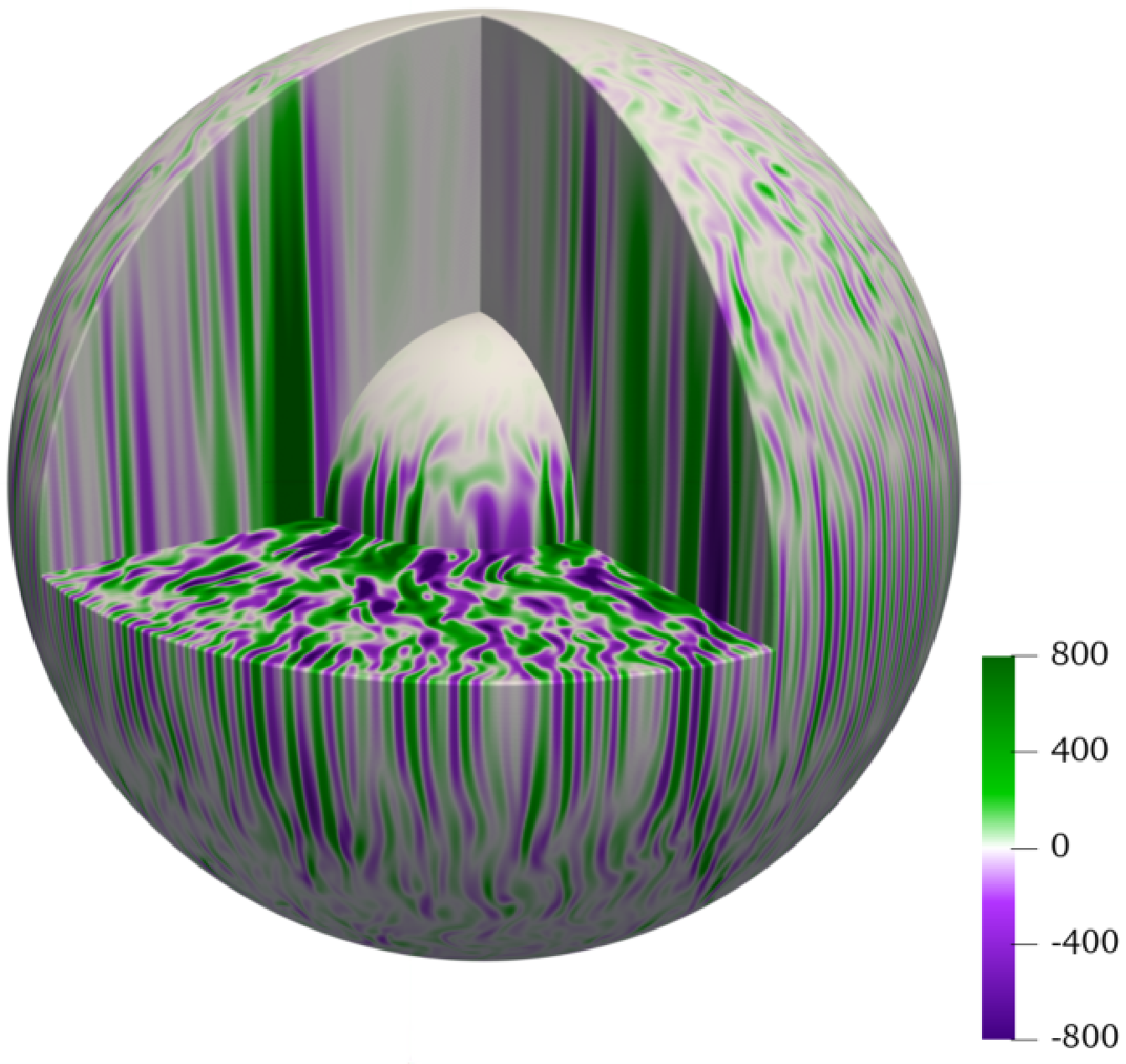}}
\subfloat[][]{\includegraphics[width=0.45\textwidth]{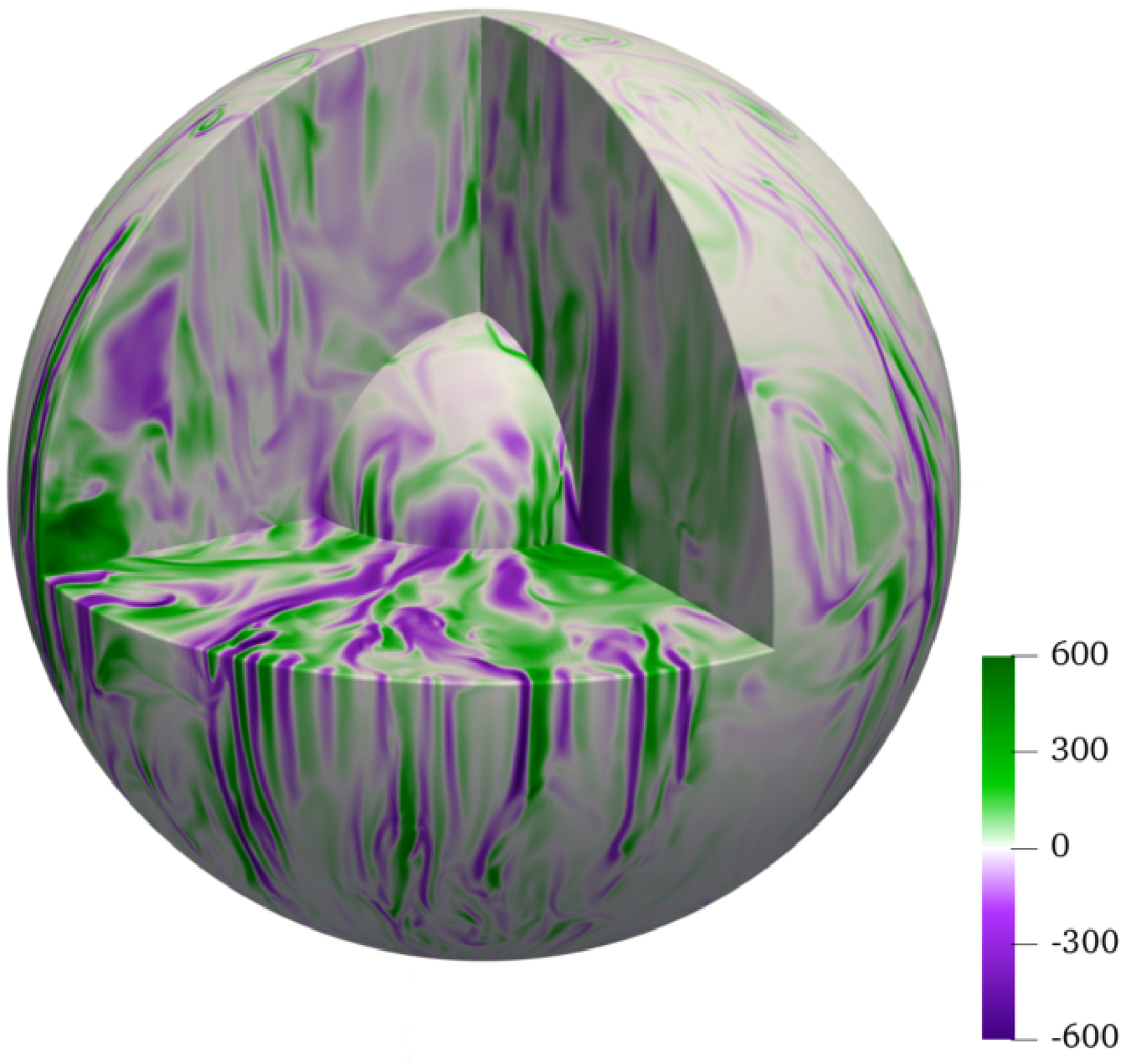}}\\
\subfloat[][]{\includegraphics[width=0.45\textwidth]{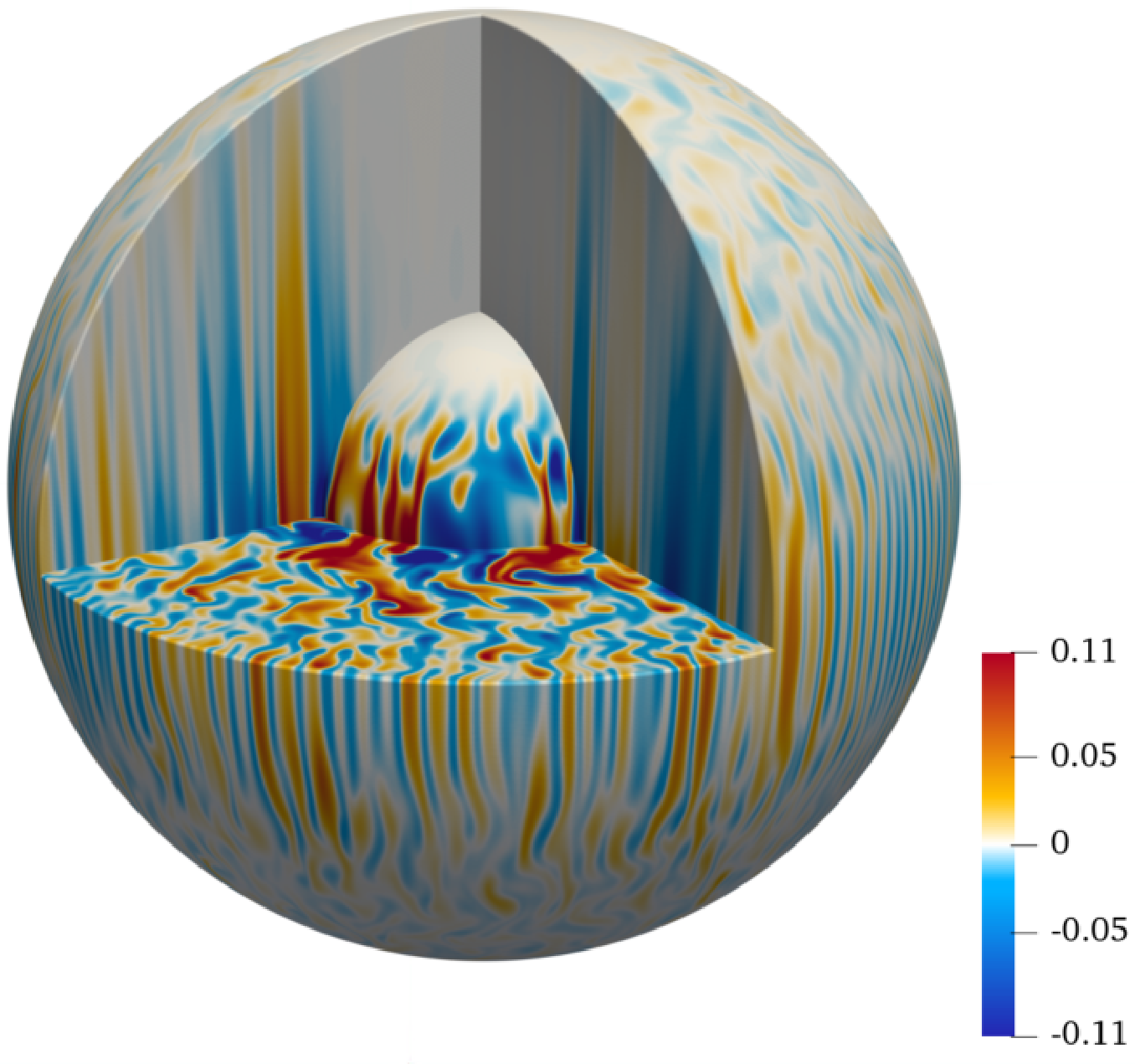}}
\subfloat[][]{\includegraphics[width=0.45\textwidth]{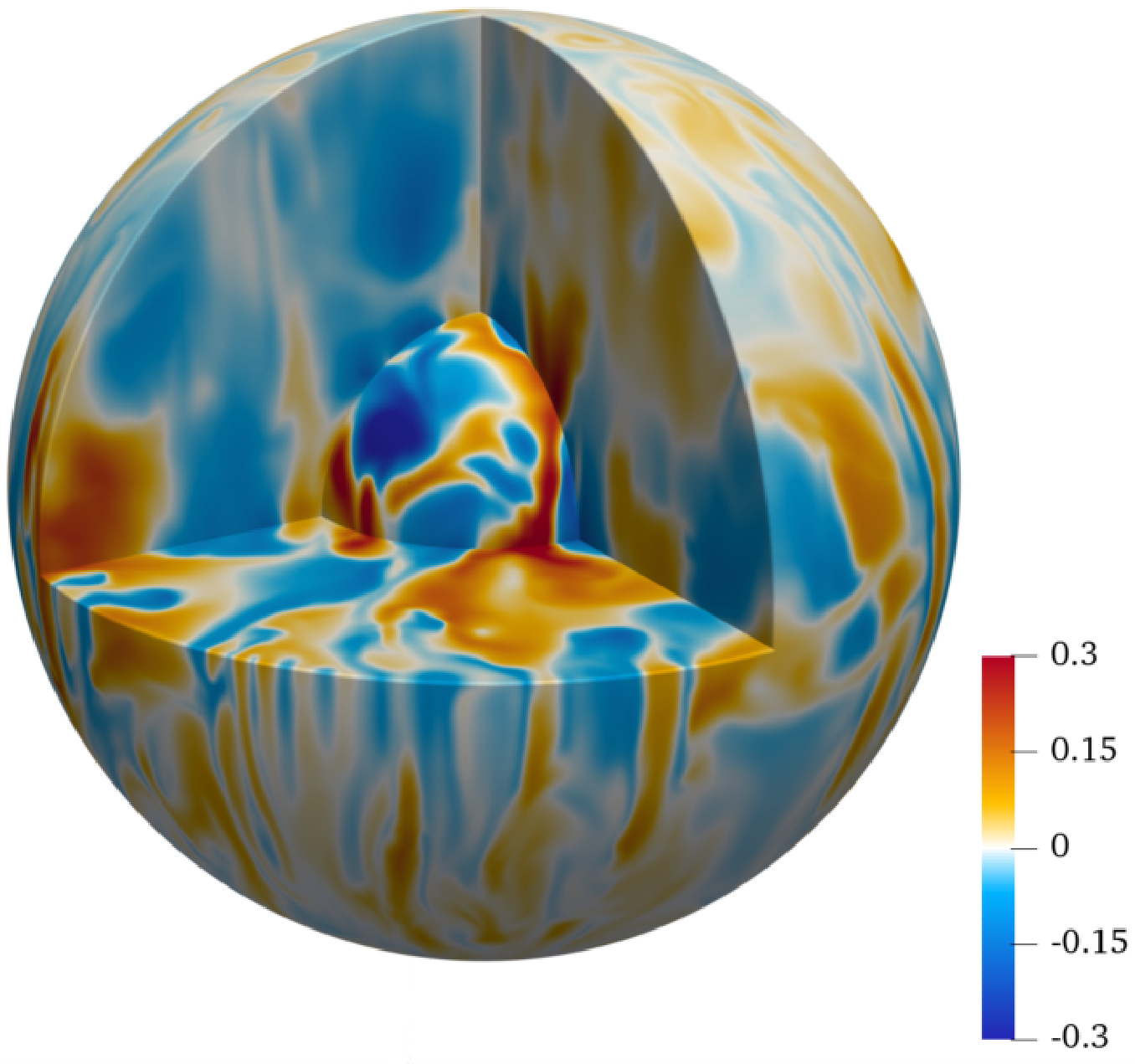}}
\caption{Visualisations of (a,b) radial velocity and (c,d) fluctuating temperature for (a,c) a non-magnetic case and (b,d) a dynamo case. For both cases $Ek=3\times 10^{-6}$ and $Ra=6.89\times 10^8$ ($\Rat = 29.8$).
%The inner and outer spherical shells in this visualisation are at $r=0.577$ and $r=1.5$, respectively. These visualisations were rendered using ParaView \protect\citep{jA05b}. 
}%\textcolor{blue}{I changed the case since this one has less ringing...}}
\label{F:slices}
\end{center}
\end{figure*}

In this section we compare various length scales for both the dynamo and non-magnetic cases. 
Figure \ref{F:slices} shows visualisations of both the velocity field and the fluctuating temperature field. As noted in previous work \citep[e.g.][]{rY16,tS19}, the dynamo cases can form length scales that are larger than those observed in comparison to the non-magnetic cases, although small length scales are still present in the dynamo cases. For example, localised small-scale fluid structures with significant flow speeds can be seen in figure \ref{F:slices}(b), especially near the inner and outer boundaries. This observation suggests that the dynamo simulations have a wider range of energetically relevant length scales than the non-magnetic cases.  Note also the general lack of axial alignment in the dynamo cases, which likely results from the Lorentz force breaking the balance between the Coriolis force and the pressure gradient force.

\begin{figure*}
\begin{center}
     \subfloat[][]{\includegraphics[width=0.45\textwidth]{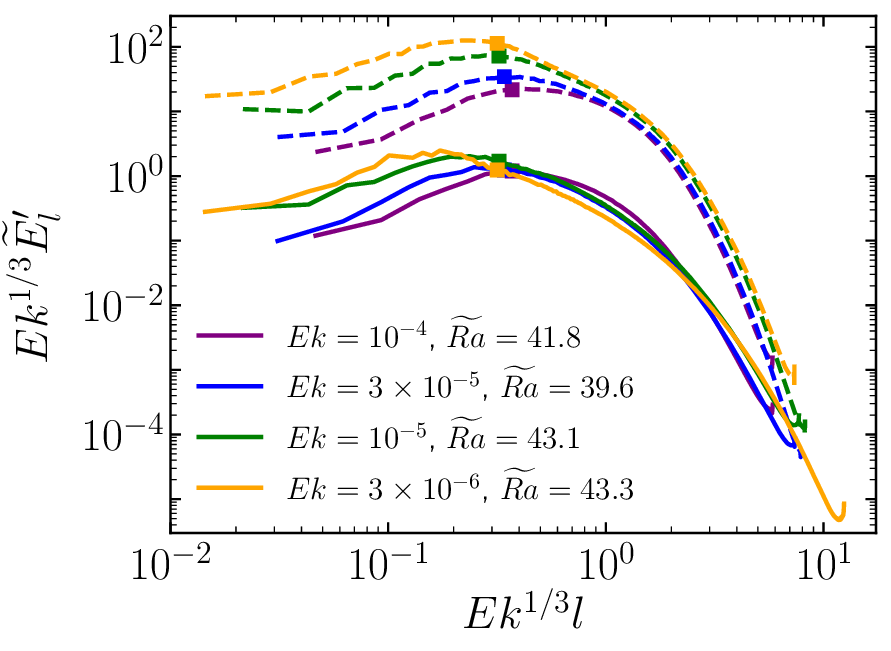}} \qquad
     \subfloat[][]{\includegraphics[width=0.45\textwidth]{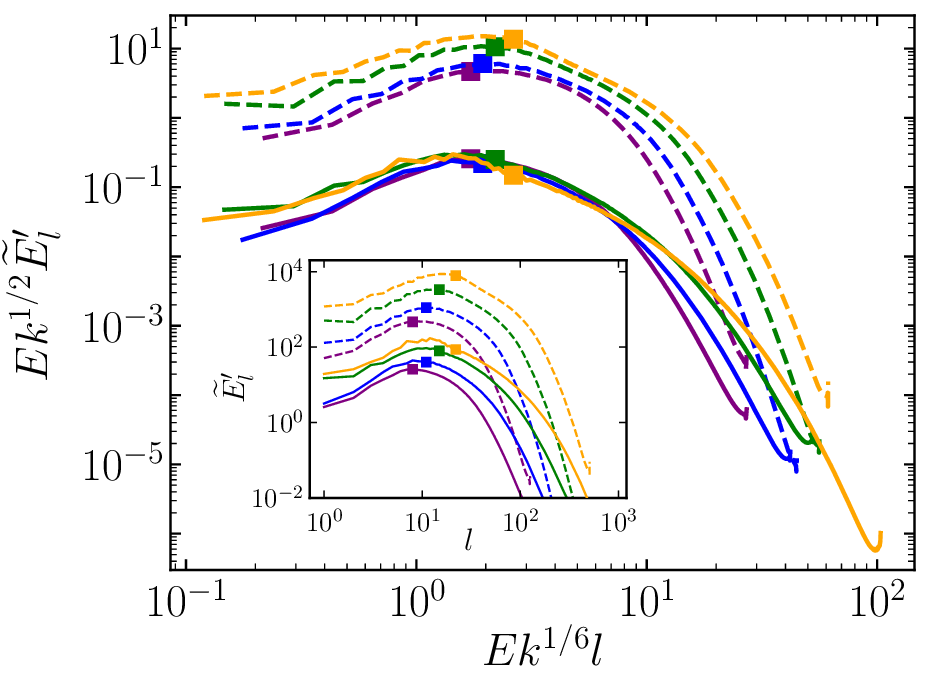}}\\
     \subfloat[][]{\includegraphics[width=0.45\textwidth]{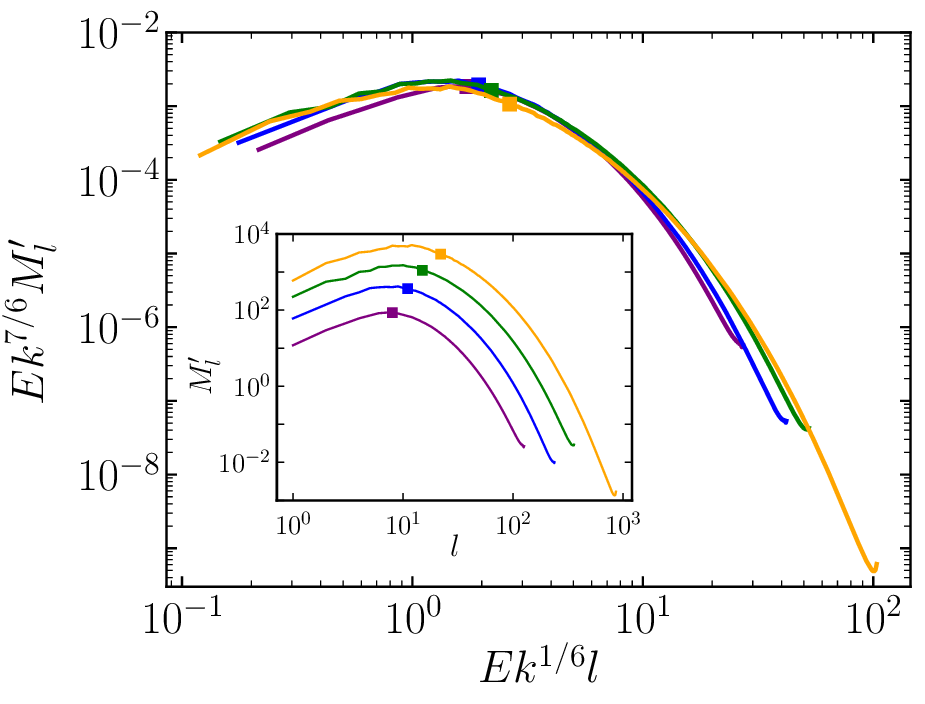}} \qquad
     \subfloat[][]{\includegraphics[width=0.45\textwidth]{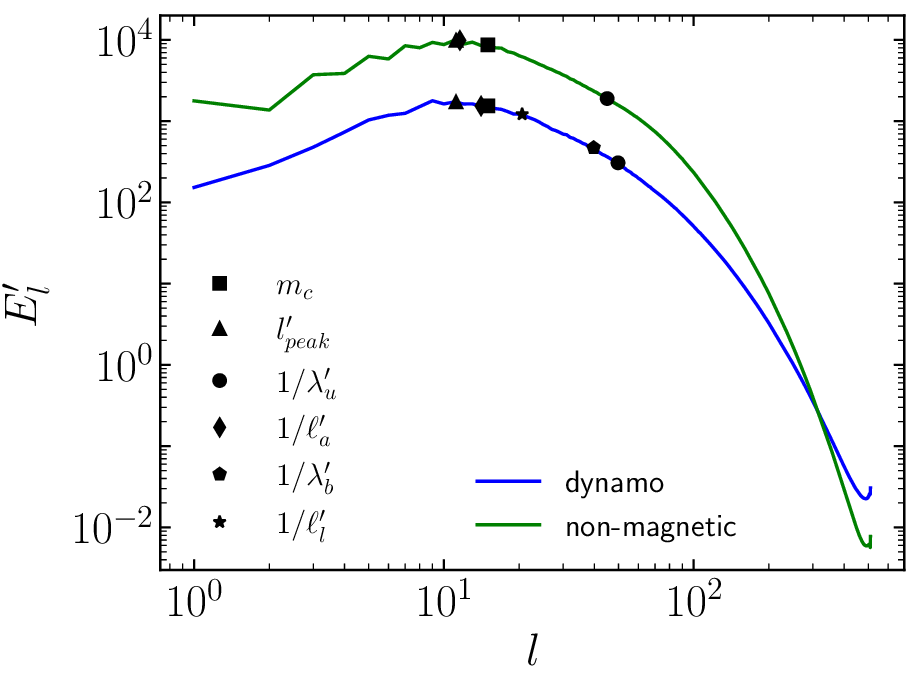}}
	 \caption{Kinetic and magnetic energy spectra for select cases, where $\widetilde{E}^{\prime}_{l}=E^{\prime}_{l}$ for the non-magnetic cases and $\widetilde{E}^{\prime}_{l}=0.1 E^{\prime}_{l}$ for the dynamo cases. (a) Rescaled kinetic energy spectra using $Ek^{1/3}$ to rescale the degree $l$. (b) Rescaled kinetic energy spectra using $Ek^{1/6}$ to rescale the degree $l$. (c) Rescaled magnetic energy spectra. (d) Kinetic energy spectra for the case $Ek=10^{-5}$, $Ra=3\times 10^{8}$ ($\Rat = 64.6$)  with different length scales shown for reference. Squares show the critical azimuthal degree $m_c$ for the onset of convection for the Ekman number of the corresponding color. The insets in (b) and (d) show the kinetic and magnetic energy spectra without rescaling, respectively.}
     \label{F:ke_spectra}
\end{center}
\end{figure*}

In order to compare how the length scales in the velocity field vary with Ekman number, figure \ref{F:ke_spectra}(a,b) shows the rescaled kinetic energy spectra at fixed reduced Rayleigh number ($\Rat \sim 40$) for both dynamo and non-magnetic cases. Figure \ref{F:ke_spectra}(a) shows the spectra rescaled according to QG theory: the spherical harmonic degree is rescaled by $Ek^{1/3}$, which roughly corresponds to rescaling the wavenumber; and the scaling for the vertical axis is chosen such that the product of the scaling of the horizontal and vertical axes gives $Ek^{2/3}$, which rescales the total kinetic energy to be order unity upon noting the scaling for the flow speeds discussed in section \ref{S:speed}.
We find excellent collapse of the spectra for the region $Ek^{1/3} l \gtrsim O(1)$, whereas a systematic deviation from this trend is evident for $Ek^{1/3} l \lesssim O(1)$, i.e.~for large length scales. That the QG scaling is effective for such a large number of spherical harmonic degrees implies that a broad range of length scales in the flow field are governed by these QG scalings. 
To investigate the scaling of large length scales (small degrees), we show the spectra rescaled by $Ek^{1/6}$ in figure \ref{F:ke_spectra}(b). For the non-magnetic cases, there is a trend in which the smaller Ekman number simulations are characterised by larger rescaled energy in comparison to the simulations with larger Ekman number; this behaviour is likely a result of the trend observed in the rescaled flow speeds.

The linear asymptotic theory of rapidly rotating convection shows that, in addition to the $Ek^{1/3}$ `fast' scale, a `slow' radial scale is also present with an asymptotic scaling of $Ek^{2/9}$ and $Ek^{1/6}$ for the spherical shell and full sphere geometries, respectively \citep{cJ00,eD04}. Thus, it might be anticipated that different regions of the spectra should be characterised by different asymptotic scalings. For our investigated range of parameters there is very little difference in these two scaling exponents so that $Ek^{1/6}$ was used here for simplicity. As shown later, this larger length scale may be pertinent to the generation of the magnetic field. 

The scaling of the magnetic energy spectra is shown in figure \ref{F:ke_spectra}(c). Rescaling the spherical harmonic degree $l$ by $Ek^{1/6}$ and the amplitude of the spectra by $Ek^{7/6}$ is found to collapse the entire range of the spectra. This uniform collapse for the scales present in the magnetic field should be contrasted with the kinetic energy spectra where different regions of the spectra follow different Ekman number dependence. These spectra suggest that the magnetic field does not develop a clear $Ek^{1/3}$ length scale over our investigated range of parameters. 

Figure \ref{F:ke_spectra}(d) shows the kinetic energy spectra for both a dynamo and non-magnetic case at $Ek=10^{-5}$, $Ra=3\times 10^{8}$. Also shown are various length scales plotted on the kinetic energy spectra at a value of $l$ that roughly corresponds to the given length scale. From the dynamo case, it can be seen that the peak length scale is the largest length scale, followed by the advection length scale, and the critical onset length scale. The two smallest length scales are the magnetic and viscous dissipation length scales, with the viscous dissipation length scale being the smaller of the two. The ordering of the length scales for the non-magnetic case is similar to that of the dynamo case. The asymptotic scaling of these various length scales will be explored next.

\begin{figure*}
 \begin{center}
     \subfloat[][]{\includegraphics[width=0.5\textwidth]{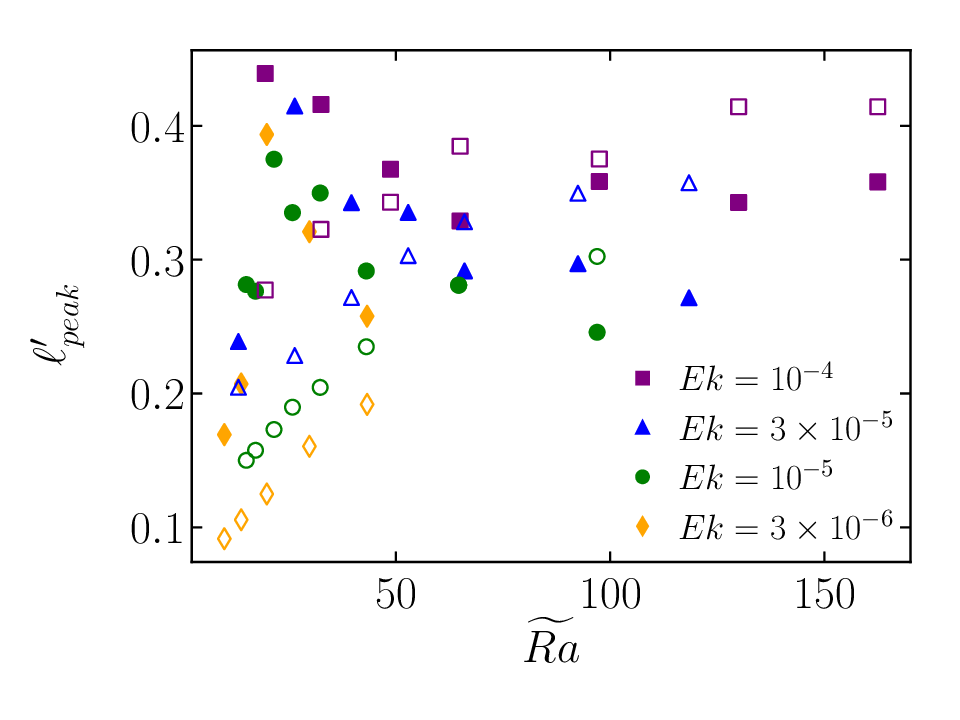}}
     \subfloat[][]{\includegraphics[width=0.5\textwidth]{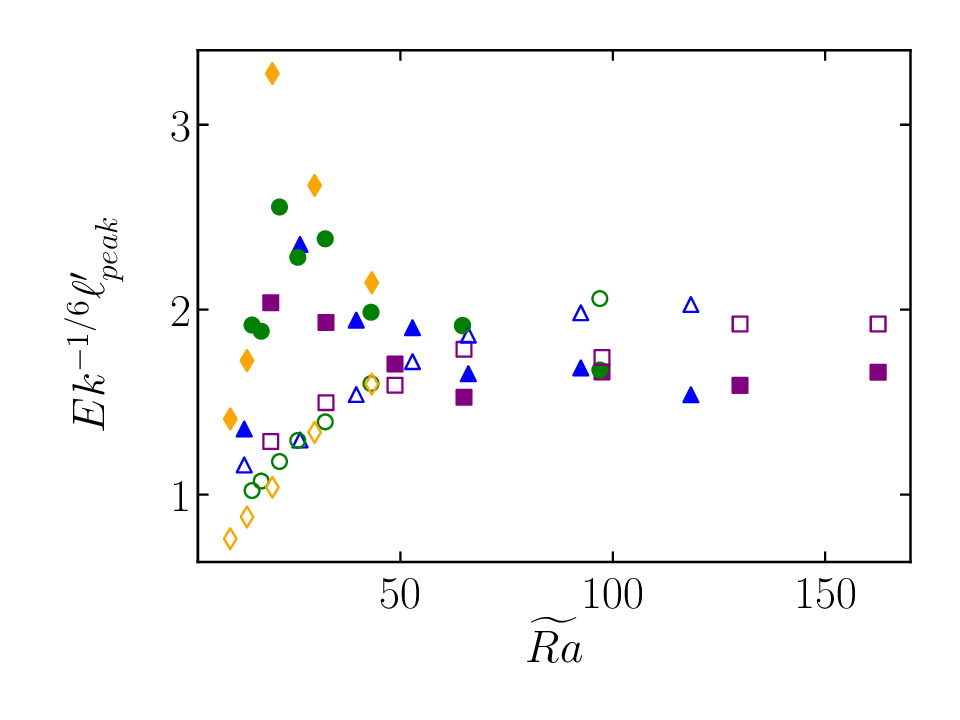}}\\
%     \subfloat[][]{\includegraphics[width=0.5\textwidth]{figures/ls_peak_Ro.eps}}
%     \subfloat[][]{\includegraphics[width=0.5\textwidth]{figures/ls_peak_Re_rescaled_half.eps}}
	 \caption{Length scales computed from the peak in the fluctuating spherical harmonic kinetic energy spectra. (a) Peak length scale; (b) rescaled peak length scale.
	  %(c) peak length scale versus the Rossby number, $Ro=EkRe_c$ (scalings of $\ell_{peak}\sim Ro^{1/2}$ and $\ell_{peak}\sim Ro^{1/4}$ are shown for reference); (d) rescaled peak length scale versus the reduced convective Reynolds number ($\widetilde{Re}_c = Ek^{1/3}Re_c$). 
	  The filled symbols denote dynamo cases and the open symbols denote non-magnetic cases. }
     \label{F:lpeak}
\end{center}
\end{figure*}

The length scale corresponding to the peak of the kinetic energy spectra, $\ell^{\prime}_{peak}$, is given in figure \ref{F:lpeak}. There is a significant difference between the peak length scale for the dynamo and non-magnetic cases at $\Rat \sim 20$, where the dynamo cases have a much larger length scale than the non-magnetic cases. The reason for this might be that the ratio of the Lorentz force to the viscous force peaks at this reduced Rayleigh number, so that the influence of the magnetic field is strong at $\Rat \sim 20$. For the $Ek=3\times 10^{-6}$ cases, the peak length scale for the dynamo cases is always larger than the peak length scale of the corresponding non-magnetic case. For $Ek=10^{-4}$, $Ek=3\times 10^{-5}$, and $Ek=10^{-5}$, the dynamo cases have the larger length scale for $\Rat \lesssim 60$, while for larger Rayleigh numbers the non-magnetic cases have a larger length scale. 
%This is due to the tendency of the length scale in the non-magnetic cases to increase with Rayleigh number. 
The non-magnetic data exhibits a systematic decrease of the length scale as the Ekman number is reduced. Figure \ref{F:lpeak}(b) shows the rescaled peak length in which a scaling of $O(Ek^{1/6})$ has been used. The dynamo cases appear to follow a similar scaling to the non-magnetic cases, although the large amount of scatter in the dynamo cases makes the scaling less clear.

\begin{figure*}
\begin{center}
\subfloat[][]{\includegraphics[width=0.45\textwidth]{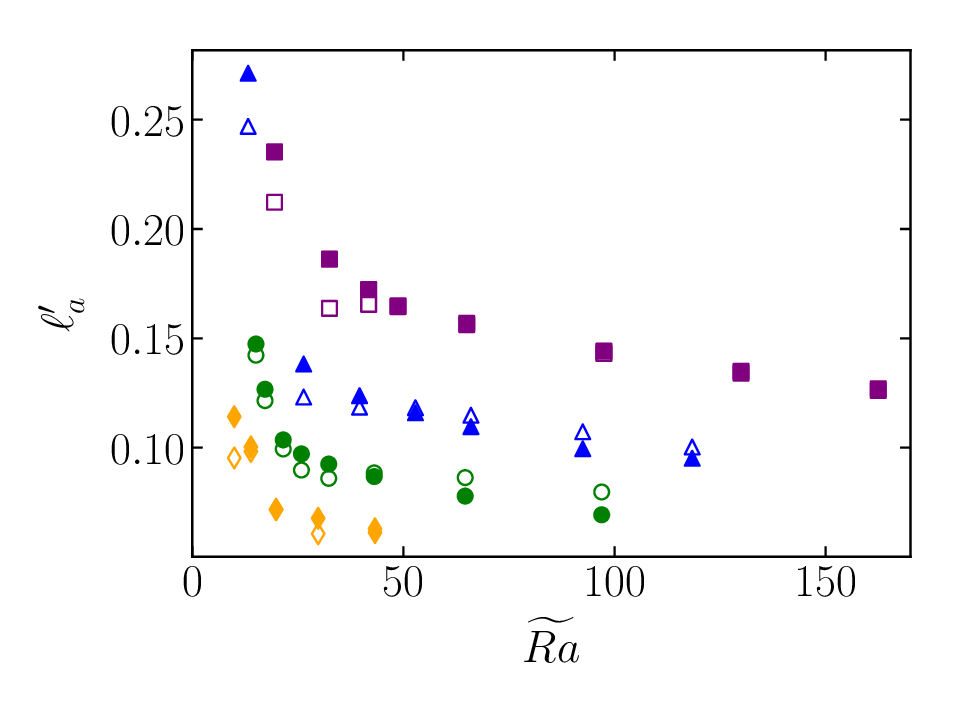}}
\subfloat[][]{\includegraphics[width=0.45\textwidth]{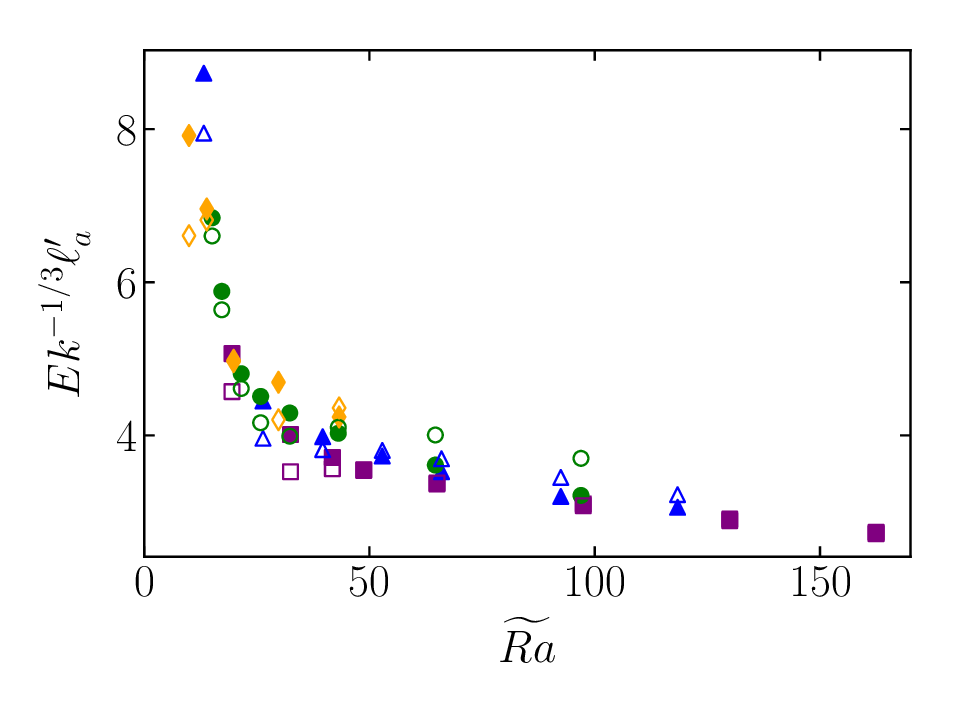}} \\
\subfloat[][]{\includegraphics[width=0.45\textwidth]{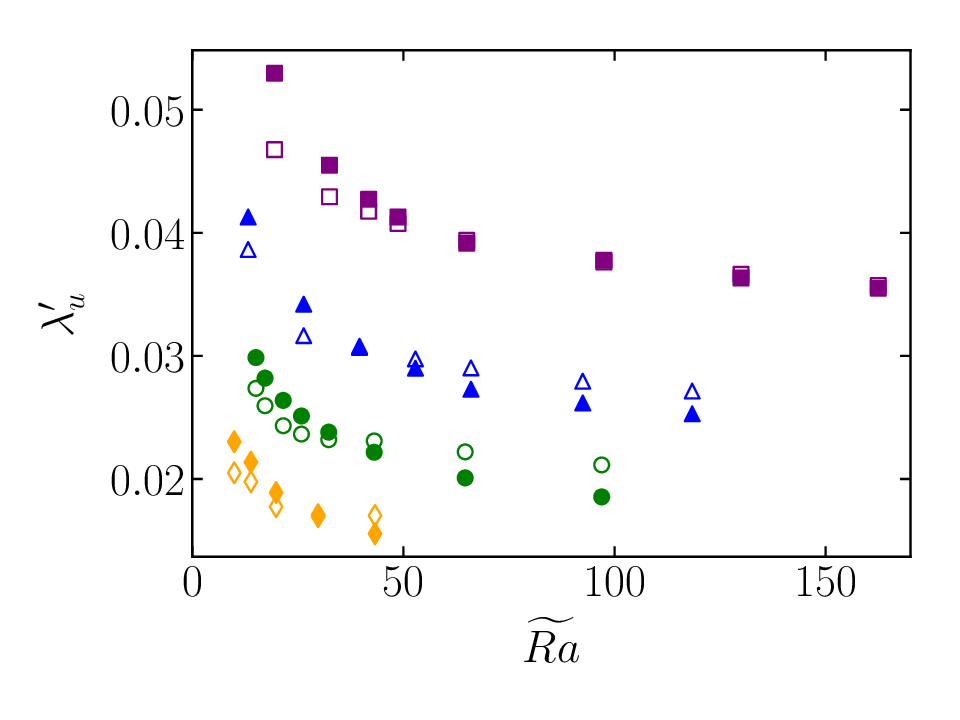}}
\subfloat[][]{\includegraphics[width=0.45\textwidth]{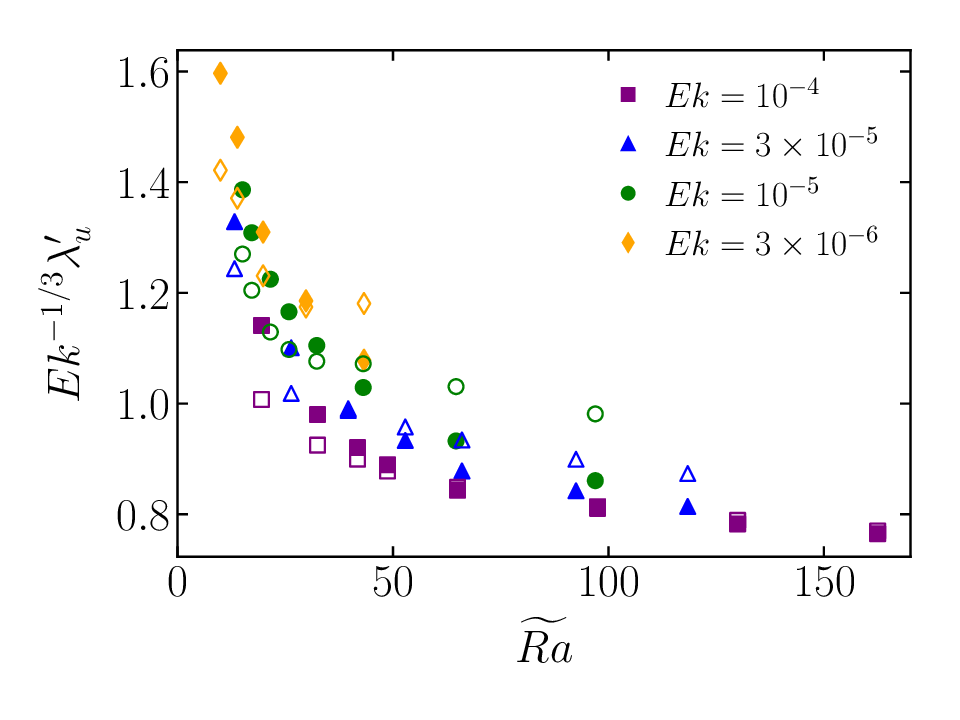}}\\
\caption{Length scales computed from the velocity field: (a) advection length scale; (b) rescaled advection length scale; (c) velocity Taylor microscale; (d) rescaled velocity Taylor microscale. The filled symbols denote dynamo cases and the unfilled symbols denote non-magnetic cases.}
\label{F:taylor_microscale}
\end{center}
\end{figure*}

One of the goals in characterising length scales is to quantify their role in various terms in the governing equations. Towards this end we define a length scale from the nonlinear advection term according to equation \eqref{E:adv}. As shown in figure \ref{F:taylor_microscale}(a), the resulting length scale is similar for both the dynamo and non-magnetic cases; this finding shows that the differences in magnitude of the advection term for these cases is not due to a difference in length scale. The asymptotically rescaled advection length scale is given in figure \ref{F:taylor_microscale}(b), showing that the QG theory holds for both sets of cases. This agreement with the theory is expected given the scalings for the forces discussed previously. We find that in contrast to the length scale computed from the peak of the kinetic energy spectra, the advection length scale generally decreases with increasing $\Rat$ for all cases considered. However, there does appear to be a slight increase in this length scale for the $Ek=10^{-5}$ non-magnetic cases at high $\Rat$.

A comparison of the velocity Taylor microscale for the dynamo and non-magnetic cases is shown in figure \ref{F:taylor_microscale}(c). The length scale for the non-magnetic cases tends to be slightly smaller than for the dynamo cases, as might be expected based on the visualisations in figure \ref{F:slices}. However, the difference in the Taylor microscale between the non-magnetic and dynamo cases is small, which might be due to viscous dissipation in the dynamo simulations being concentrated in small, localized regions. In addition, the Taylor microscale scales approximately as $O(Ek^{1/3})$ as shown in figure \ref{F:taylor_microscale}(d), although there is still a weak trend in the rescaled data for both the dynamo and non-magnetic cases. This trend might suggest a slow rate of convergence to the predicted QG scaling. Other studies have also found that the velocity field retains an $O(Ek^{1/3})$ length scale \citep{eK13b}, though we stress here that this scaling does not necessarily indicate that the viscous force is dominant since, as we have shown, the same length scale can be deduced from the advection term.

\begin{figure*}
\begin{center}
\subfloat[][]{\includegraphics[width=0.45\textwidth]{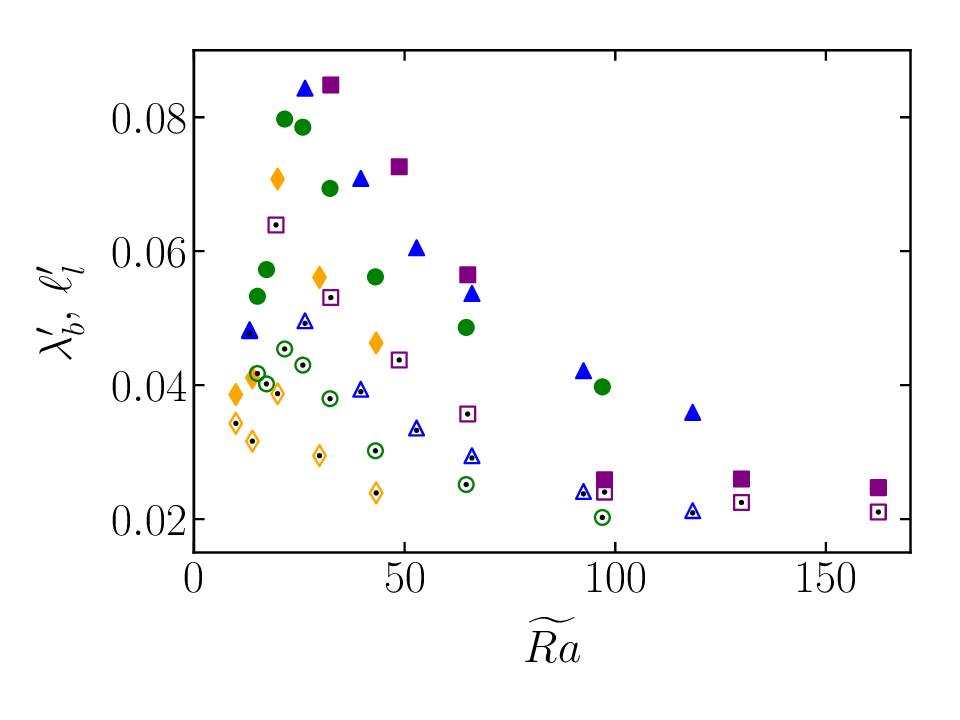}}
\subfloat[][]{\includegraphics[width=0.45\textwidth]{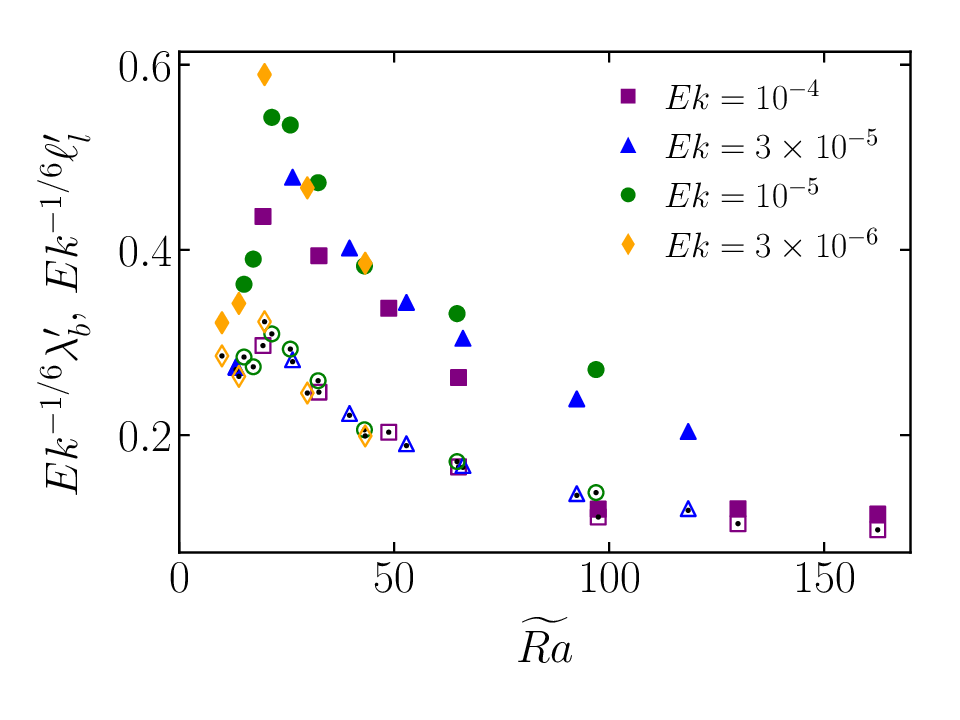}}\\
\caption{Length scales computed from the magnetic field. The filled symbols show the length scale calculated from the Lorentz force, $\ell_l^{\prime}$, and the open symbols with a central dot show the magnetic Taylor microscale, $\lambda_b^{\prime}$. (a) Magnetic length scales. (b) Rescaled magnetic length scales. }
\label{F:magnetic_ls}
\end{center}
\end{figure*}

Finally, two magnetic length scales are shown in figure \ref{F:magnetic_ls}, where both the magnetic Taylor microscale ($\lambda^{\prime}_b$) and a length scale based on the Lorentz force ($\ell^{\prime}_b$) are shown. These length scales are defined in equations \eqref{E:tay} and \eqref{E:lor}. Both length scales are comparable in magnitude over the investigated range of parameters and both follow similar trends with $\Rat$ and $Ek$. In particular, for smaller values of $\Rat$ these length scales show an increase up to $\Rat \sim 20-30$, and then a monotonic decrease is observed as $\Rat$ is increased. Figure \ref{F:magnetic_ls}(b) shows the magnetic length scales rescaled with $O(Ek^{1/6})$ and a good collapse is observed.
Previous studies have found that the length scale of the magnetic field varies according to $Rm^{-1/2}$ \citep[e.g.][]{uC04,kS15,mM20}. We can relate this scaling to the Ekman number via
\be
Rm^{-1/2} = \lb Pm Re \rb^{-1/2} = O \lb Ek^{1/6} \rb,
\ee
since $Pm = O(1)$ and $Re = O(Ek^{-1/3})$ in our cases. We note that whereas the $O(Ek^{1/6})$ scaling in the magnetic Taylor microscale is weaker than the $O(Ek^{1/3})$ scaling in the velocity Taylor microscale, the computed values for both of these length scales are comparable over the parameter regime covered here \citep[cf.][]{jA17,pD13b}.

\section{Conclusion}
\label{S:conclusion}

\begin{table*}
%\begin{adjustwidth}{-1.5cm}{}
\footnotesize
%\scriptsize
\centering
\begin{tabular}{|c|c|c|c|}
\hline
Quantity                          &                      QG theory                 &             observed ($\mathbf{B}=0$)           &             observed ($\mathbf{B}\ne0$)                   \\
\hline
$Re_c$                            &                  $O(Ek^{-1/3})$            &                $O(Ek^{-1/3})$                         &                $O(Ek^{-1/3})$                \\
\hline
$T'$                                &                   $O(Ek^{1/3})$            &                $O(Ek^{1/3})$                           &                $O(1)$                       \\
\hline
$F_c^{\prime}$               &                   $O(Ek^{-4/3})$           &               $O(Ek^{-4/3})$                           &                $O(Ek^{-4/3})$               \\
\hline
$F_p^{\prime}$               &                   $O(Ek^{-4/3})$           &               $O(Ek^{-4/3})$                           &                $O(Ek^{-4/3})$               \\
\hline
$F_b^{\prime}$               &                    $O(Ek^{-1})$             &               $O(Ek^{-1})$                              &                $O(Ek^{-4/3})$             \\
\hline
$B^{\prime}$                    &                  $O(Ek^{-1/6})$            &                          ---                                  &                 $O(Ek^{-1/6})$              \\
\hline
$F_l^{\prime}$                &                    $O(Ek^{-1})$             &                     ---                                        &                $O(Ek^{-3/2})$               \\
\hline
$F_v^{\prime}$               &                   $O(Ek^{-1})$              &               $O(Ek^{-1})$                              &                $O(Ek^{-1})$               \\
\hline 
$F_a^{\prime}$               &                    $O(Ek^{-1})$             &               $O(Ek^{-1})$                              &                $O(Ek^{-1})$               \\
\hline
$\lambda_u^{\prime}$      &                    $O(Ek^{1/3})$          &                $O(Ek^{1/3})$                           &                 $O(Ek^{1/3})$              \\
\hline
$\ell_a^{\prime}$             &                    $O(Ek^{1/3})$          &                $O(Ek^{1/3})$                           &                 $O(Ek^{1/3})$              \\
\hline
$\ell_{peak}$                    &                    $O(Ek^{1/3})$          &                $O(Ek^{1/6})$                           &                 $O(Ek^{1/6})$              \\
\hline
$\lambda_b^{\prime}$      &                    $O(Ek^{1/3})$          &                          ---                                   &                 $O(Ek^{1/6})$              \\
\hline
$\ell_l^{\prime}$               &                     $O(Ek^{1/3})$         &                          ---                                  &                 $O(Ek^{1/6})$              \\
\hline
$\varepsilon_u$                 &                    $O(Ek^{-4/3})$         &               $O(Ek^{-4/3})$                          &                $O(Ek^{-4/3})$             \\
\hline
$\varepsilon_b$                 &                     $O(Ek^{-4/3})$        &                       ---                                      &                $O(Ek^{-5/3})$             \\
\hline
\end{tabular}
\caption{Summary of asymptotic scalings for various quantities. Second column: predictions from QG plane layer theory \citep{mC15b}. Third column: observed non-magnetic scalings. Fourth column: observed dynamo scalings. These scalings assume the governing equations have been non-dimensionalised using the shell depth and the kinematic viscosity. The QG theory prediction for $B'$ assumes $Pm$ is fixed.}
\label{T:scalings}
%\end{adjustwidth}
\end{table*}

Understanding the dynamics of Earth's outer core and the interiors of other planets and stars requires the extrapolation of model output by many orders of magnitude in parameter space. The identification of asymptotic behaviour in model output is paramount for this extrapolation process. A key component of this process is the identification of the asymptotically small (or large) parameter. In non-magnetic rotating convection this parameter is the Ekman number, $Ek$, and all other small parameters present, including the Rossby number, can be directly tied to $Ek$ via so-called distinguished limits. The dynamics of rapidly rotating convection is QG; a rigorous derivation of the nonlinear three-dimensional QG model is possible in planar \citep{mS06}, and annular geometries with small radial length scales \citep{mC13}. Recent work shows that the dominant scalings characterising the small-scale dynamics in both planar and spherical geometries are identical when the magnetic field is either not sufficiently strong to disrupt the geostrophic balance \citep{mY22,mY22b} or is not present at all \citep{jN24}. Given that the force balance present within the outer core is not directly observable, it is important to determine how the magnetic field influences the dynamics and resulting scaling behaviour as it becomes increasingly strong. Towards this end, we have used a set of numerical models \textcolor{black}{with $Pr=1$, $Pm=2$ (with a subset of simulations varying $Pm$ from $0.2$ to $5$), $Ek \in \{3\times 10^{-6}, 10^{-5}, 3\times 10^{-5}, 10^{-4} \}$ and $Ra$ up to $50$ times supercritical} to determine how the dynamics depend on the Ekman number, and compared the results with equivalent simulations in which no magnetic field is present. Our findings, given in terms of the asymptotic scalings, are summarized in table \ref{T:scalings}. These results help to explain the behaviour of the forces studied in prior work \citep[e.g.][]{kS15,rY16}, and therefore provide a quantitative measure for the degree of asymptoticness that is obtained in any given simulation.

The dynamo cases exhibit a MAC force balance for sufficiently large values of $Pm$, with the Coriolis force, pressure force, buoyancy force, and Lorentz force entering at leading order. \textcolor{black}{This large-scale balance has been observed in other direct numerical simulations that managed to perform a limited number of simulations at lower $Ek$ and $Pm$ than our model suite \citep{nS17, yL25}. In our simulations,} the Lorentz force was observed to follow a stronger scaling with the Ekman number than the Coriolis force and buoyancy force. This finding suggests that either the Lorentz force becomes asymptotically larger than the Coriolis force as the Ekman number is decreased, in which case the Lorentz force would have to be balanced by the pressure gradient force, or that the scaling must change at smaller Ekman number beyond the current reach of simulations. The buoyancy force is found to be approximately the same order as the Coriolis force as $\Rat$ is increased, which, when combined with the approximately constant fluctuating temperature, leads to a simple relation for the Reynolds number. Notably, this scaling relationship, $Re_c \sim Ra Ek/Pr$, is equivalent to that derived from the CIA balance \citep[e.g.][]{jmA20}, despite the fact that the balances used to derive the relationship occur at different asymptotic orders. The order unity temperature fluctuation in the dynamos implies that the buoyancy force is the same asymptotic order as the Coriolis force whereas the buoyancy force for the non-magnetic cases is asymptotically smaller than the Coriolis force; this asymptotic difference in buoyancy force is the most notable difference between the dynamo and non-magnetic cases. We have argued that this difference in asymptotic order likely arises from the ohmic dissipation term in the dissipation equation.

The magnetic Prandtl number plays a critical role in dynamos since, \textcolor{black}{for fixed values of $\Rat$ and $Ek$, it can be used to control the relative size of the Lorentz force and therefore the observed force balance.} Previous work has often used the strategy of employing as small a value of $Pm$ as can be used to generate dynamo action, such that the Lorentz force might play a secondary role. The majority of the simulations used in the present work have focused on a relatively large value of $Pm=2$, since if changes in the asymptotic behaviour occur due to the influence of magnetic fields, then such changes should be more readily observable with strong magnetic fields. For sufficiently small $Pm$, studies of plane layer dynamos find that the buoyancy force and Lorentz force enter at approximately the same order as the viscous force \citep{mY22}, whereas in the present study at constant $Pm$ both the buoyancy force and Lorentz force enter at an asymptotic order at least as large as the Coriolis force. It is possible that these two forces can enter at any asymptotic order between the viscous force and Coriolis force for an appropriate choice of $Pm$. The advection term, viscous force, and Coriolis force appear to follow the same asymptotic scaling at the fixed value of $Pm$ used in this paper as was found in the low $Pm$ values of \cite{mY22}, which suggests these terms do not depend asymptotically on $Pm$, although they are not independent of $Pm$. \textcolor{black}{Future work is needed to understand the dynamics in the dual limit $(Ek,Pm) \rightarrow 0$.}

The kinetic energy spectra computed from the non-axisymmetric motions show that the simulations are characterised by a broad range of length scales in which different regions of the spectra exhibit different asymptotic dependencies on the Ekman number. Similar behaviour was observed in non-magnetic spherical convection with stress-free boundary conditions \citep{jN24}. The largest length scales, as computed from the peak of the kinetic energy spectra, show an approximate scaling of $O(Ek^{1/6})$, whereas scales smaller than this are well characterised by the $O(Ek^{1/3})$ scaling. The flow speeds, in agreement with QG theory, follow a $Re = O\lb Ek^{-1/3} \rb$ scaling. Prior work has suggested that the presence of $O(Ek^{1/3})$ length scales in simulations is an indication that viscosity plays a dominant role in the dynamics \citep{eK13b}. We stress that this conclusion is not generally true. Rather, this scaling arises from the asymptotic nature of many different terms in the governing equations and does not in itself indicate a dominant role of viscosity. Indeed, as shown here, a length scale derived from the advection term also exhibits this same asymptotic scaling. 

Length scales computed from the magnetic field are approximately characterised by a $O(Ek^{1/6})$ scaling, i.e.~substantially weaker than the dominant $O(Ek^{1/3})$ scaling observed in the velocity field. As discussed in prior work, the magnetic induction equation can be used to deduce a ohmic dissipation length scale that follows a $Rm^{-1/2}$ scaling \citep[e.g.][]{uC04}. Here we have shown that this scaling is equivalent to an asymptotic dependence of $O(Ek^{1/6})$ for $Pm = O(1)$. Further work is necessary to determine the reason for the differences in length scales in the magnetic and velocity fields.

Similar to other studies \citep[e.g.][]{eD16,mM20}, we find that increasing the magnetic Prandtl number can greatly increase the Lorentz force, and we also find that the buoyancy force can increase with $Pm$. This seems to be in contradiction to diffusion-free scaling arguments, which assume that the force balance is independent of $Pm$ \citep[e.g.][]{pD13b}. The question remains as to whether dynamos are independent of magnetic diffusivity at more extreme parameter values, which could be tested directly by varying $Pm$ for low Ekman number high Rayleigh number simulations. If so, it would be interesting to find at what parameter values simulations become independent of $Pm$, which could help guide future investigations.

It is often assumed that advection plays no role in the geodynamo because the Rossby number in the outer core is small. However, this argument is oversimplified given what is known about the perturbative dynamics of rapidly rotating convection; while both inertia and advection are small relative to the leading order geostrophic balance, they nevertheless play crucial roles in the dynamics of geostrophic convective turbulence, even for flows with an asymptotically small Rossby number \citep[e.g.][]{kJ12}. 
Moreover, a plethora of waves and time-dependent flows are thought to be important in the core \citep[e.g.][]{cF23}. The simulations analysed here confirm that advection is small relative to the leading order forces, though we find that it is comparable in magnitude to, and shows the same asymptotic dependence as, both inertia and the viscous force. This finding suggests that the temporal dynamics are characterised by an $O(Ek^{2/3})$ timescale (in our non-dimensional units), which is equivalent to the convective Rossby wave timescale from asymptotic theory \citep[][]{pR68}, and found to be the dominant timescale in prior studies of turbulent rotating convection \citep{tO25}. This result is surprising in light of the observed force balance (MAC). On the other hand, the prevalence of the $Ek^{1/3}$ in the velocity field essentially requires this to be the dominant inertial timescale.

\cite{jA17} and related subsequent work \citep[e.g.][]{tS19,tS21} utilised spherical harmonic power spectra of the forces in the governing equations in an effort to better understand the relationship between force balances and length scales in both non-magnetic rotating convection and rotating convection-driven dynamos. By basing their analysis on the spectral representation of the forces, as opposed to physical space, they conclude that dynamos with sufficiently strong magnetic field are geostrophically balanced to leading order on large length scales (small spherical harmonic degrees), and the Lorentz force can act to perturb this balance at some intermediate length scale. The authors refer to such dynamos as `QG-MAC', even though the physical space force balance can be MAC. One of the reasons for the discrepancy between how our force balances are characterised in physical space and their corresponding spectral analysis is due to the behaviour of the force spectra, which are rarely characterised by well-defined peaks, implying that a large number of spherical harmonic degrees contribute to the dynamics. Indeed, it is the sum over all of these degrees that leads to the correct magnitude of the forces, and therefore the force balance, in physical space. It is also important to note that the axisymmetric component of the forces is not removed from their spectra, even though the axisymmetric and non-axisymmetric dynamics can be characterised by distinct force balances, as shown in \cite{mC21} and the present work. The axisymmetric component of the forces tends to be dominated by even spherical harmonic degrees and can have a strong influence on the spectra both in terms of their amplitude and shape, as indicated by the `sawtooth' pattern in their spectra. Furthermore, \cite{rT23} showed that while the spherical harmonic representation of the forces may show distinct `crossings' in spectral space, which are used as proxies for length scales, the spherical harmonic representation of the curl of the forces (i.e.~using the vorticity equation) lacks such spectral crossings, which represents a possible issue given that dynamo scaling laws often rely on assuming balances in the vorticity equation.

\section*{Acknowledgements}
This work was supported by the Geophysics Program at the National Science Foundation (NSF) through grant numbers EAR-1945270 and EAR-2201595. CD acknowledges support from the Natural Environment Research Council of the UK under grant number NE/V010867/1. Simulations were carried out on the Alpine supercomputer at the University of Colorado Boulder, the Anvil supercomputer at Purdue University and Stampede2 at the Texas Advanced Computing Center (TACC). Alpine is jointly funded by the University of Colorado Boulder, the University of Colorado Anschutz, and Colorado State University and with support from NSF grants OAC-2201538 and OAC-2322260. Anvil and Stampede2 were made available through allocation PHY180013 from the Advanced Cyberinfrastructure Coordination Ecosystem: Services \& Support (ACCESS) program, which is supported by NSF grants 2138259, 2138286, 2138307, 2137603 and 2138296. Visualisations were made using paraview \citep{jA05b}.

%\section*{DATA AVAILABILITY}
%The data used in this study will be shared upon reasonable request to the corresponding author.

\printbibliography

\end{document}